\documentclass[11pt]{article}

\usepackage[letterpaper, margin=1in]{geometry}

\usepackage[utf8]{inputenc}
\usepackage[T1]{fontenc}
\IfFileExists{lmodern.sty}
  {\usepackage{lmodern}\usepackage{microtype}}
  {\usepackage[expansion=false]{microtype}}
\IfFileExists{babel.sty}{\usepackage[english]{babel}}{}

\usepackage{amsmath}
\usepackage{amssymb}
\usepackage{graphicx}
\usepackage{booktabs}
\usepackage{longtable}
\usepackage{array}
\usepackage{calc}
\usepackage{caption}
\usepackage{enumitem}
\setlist{itemsep=2pt, topsep=4pt, parsep=0pt}
\usepackage{setspace}
\usepackage{fancyhdr}
\usepackage{xcolor}

\definecolor{linknavy}{rgb}{0.10,0.25,0.55}
\usepackage[colorlinks=true,
            linkcolor=black,
            citecolor=black,
            urlcolor=linknavy,
            breaklinks=true]{hyperref}
\usepackage{xurl}
\DeclareUnicodeCharacter{2248}{\ensuremath{\approx}}
\DeclareUnicodeCharacter{0119}{\k{e}}
\DeclareUnicodeCharacter{00E6}{\ae}

\fancypagestyle{plain}{\fancyhf{}\fancyfoot[C]{\thepage}}

\title{\bfseries\LARGE Non-Great-Power Conflict and AI Risk}

\author{%
  Kristina Kempkey\textsuperscript{*\,\textdaggerdbl}
  \and
  Se\'an Boddy\textsuperscript{*\,\textsection}
  \and
  Catherine Ge-Wang\textsuperscript{*\,\textdagger\,\textparagraph}%
}

\date{}

\begin{document}

\maketitle
\thispagestyle{plain}

{\renewcommand{\thefootnote}{\fnsymbol{footnote}}%
 \footnotetext[1]{Future Impact Group. All three authors contributed equally.
   \textsuperscript{\textdagger}Corresponding author:
   \href{mailto:catherine.jg.wang@gmail.com}{\texttt{catherine.jg.wang@gmail.com}}.
   \textsuperscript{\textdaggerdbl}MATS; Irregular Warfare Initiative.
   \textsuperscript{\textsection}School of Computer Science and Statistics, Trinity College Dublin.
   \textsuperscript{\textparagraph}Mathematical Institute, University of Oxford.}}

\renewcommand{\thefootnote}{\arabic{footnote}}
\setcounter{footnote}{0}

\begin{abstract}
\noindent
Research on advanced AI and the risk of war has focused almost exclusively on
great power conflict, on the grounds that confrontation between nuclear-armed
adversaries poses the greatest risk of catastrophic or existential harm.
Considerably less attention has been paid to \emph{non-great-power conflict}
(NGPC): wars between non-great powers, between non-great powers and great powers, civil wars, proxy wars, and conflicts
involving nonstate actors. This paper evaluates the null hypothesis that NGPC is
much less important than great power conflict (GPC) as a source of catastrophic
risk in an era of increasingly capable AI, against the alternative that it is
within an order of magnitude of GPC in importance. We assess three
sub-hypotheses: that NGPC increases the likelihood of great power conflict; that
it increases the expected harm from catastrophic terrorism; and that it
increases the expected harm from loss of control over advanced AI systems. For
each, we construct a causal model linking NGPC to the risk outcome, decompose
that model into a parameterized multiplicative risk model where the evidence
permits, and assess the parameters qualitatively using literature review,
historical case studies, and the PHIA probability yardstick. We find the null poorly supported for H1 and H2, and identify H3 as a priority for further work rather than a settled finding. We do not estimate GPC risk directly and therefore make no quantitative comparison. Our claim is that the pathways from NGPC to catastrophic risk are numerous, mutually reinforcing, and sufficiently underexamined that the field's default assumption of a large gap should not be taken for granted, and that further investigation, at a minimum into the cause area's tractability, is warranted. We also identify five intermediate
variables that recur across the pathways---information environment quality,
decision-making timeline compression, great power threat perception, capability
diffusion, and norm erosion---and argue that these shared nodes are the
highest-priority targets for further investigation and intervention.
\end{abstract}

\vspace{1em}

\section*{Introduction}
\addcontentsline{toc}{section}{Introduction}\label{intro}

There has been growing interest in the governance and national security implications of advanced AI. In particular, since Leopold Aschenbrenner's \emph{Situational Awareness,} there has been an increasing amount of research done on the relationship between advanced AI and the risk of war. Existing literature almost ubiquitously focuses on \emph{great power conflict,} because those conflicts are seen as the main risk of catastrophic or existential harm. As a result, there has been significantly less attention paid to other forms of conflict, including civil wars and proxy wars, or conflicts involving nonstate actors in general. We call those other forms of conflict \emph{non-great-power conflict}.

This paper evaluates the null hypothesis that non-great-power conflict (NGPC) is much less important than great power conflict (GPC) as a source of catastrophic risk in the era of increasingly capable AI. Throughout, our concern is with how advanced AI reshapes the channels through which conflict generates catastrophic risk, accelerating escalation between great powers, amplifying the harm from catastrophic terrorism, and, contributing to loss of control over AI systems themselves. The alternative hypothesis is that NGPC is within an order of magnitude of the importance of GPC. We do not estimate GPC risk. The order-of-magnitude framing sets the decision-relevant threshold that motivates the question, and our contribution is to assess whether the pathways from NGPC to catastrophic risk are strong enough to make that threshold plausible, rather than to measure the ratio. If they are, then further investigation, at a minimum into the cause area's tractability, is warranted.

We assess the importance of NGPC according to three sub-hypotheses:

\begin{itemize}
\item
  \begin{quote}
  \textbf{H1: NGPC increases the likelihood of great power conflict;}
  \end{quote}
\item
  \begin{quote}
  \textbf{H2: NGPC increases the expected harm from catastrophic terrorism;}
  \end{quote}
\item
  \begin{quote}
  \textbf{H3: NGPC increases the expected harm from loss of control over advanced AI systems.}
  \end{quote}
\end{itemize}

\textbf{Method.} We proceed in three steps: a causal model for each sub-hypothesis, a parameterized decomposition of that model where the evidence permits, and a qualitative assessment of the resulting parameters. Section 1 sets out the approach in full.

We identify shared intermediate variables across the risk pathways: \emph{information environment quality}, \emph{decision-making timeline compression}, \emph{great power threat perception}, \emph{capability diffusion}, and \emph{norm erosion}. A variable that appears across multiple risk pathways is, \emph{ceteris paribus}, a higher-priority target for investigation and intervention, because interventions that shift them would affect multiple risk pathways at once. We do not model interaction effects between pathways. This cuts both ways. Pathways that reinforce one another mean our independent treatment understates the aggregate effect, while pathways driven by a common factor, particularly AI capability diffusion, risk being counted more than once. We flag the shared drivers explicitly rather than attempting to net the two effects against each other.

\subsection*{H1: NGPC Increases the Likelihood of Great Power Conflict}\label{h1-ngpc-increases-the-likelihood-of-great-power-conflict}

We construct two formal risk models.

Model A is event-centric:

P(GPC\textbar NGPC) = NGPC Rate $\times$ GP Involvement Rate $\times$ Escalation Rate $\times$ (1 $-$ De-escalation Rate).

Model B is system-centric:

P(GPC\textbar NGPC) = NGPC Instability Effect $\times$ Baseline GP Crisis Rate $\times$ Crisis-to-GPC Escalation Rate.

We investigate three pathways:

\begin{enumerate}
\def\labelenumi{\arabic{enumi}.}
\item
  \begin{quote}
  \textbf{Proxy escalation and alliance entrapment.} Contemporary high-risk cases include the South China Sea, India-Pakistan (which both have opposing great power patrons for the first time), and Sudan (six external state actors, Russian and American naval forces in operational proximity). The Cold War base rate of zero reflects not low risk but successful, and sometimes barely successful, management of continuously high risk --- resting on bilateral communication, hours-long decision timelines, principal-agent slack, and mutual recognition of nuclear redlines.
  \end{quote}

  \begin{enumerate}
  \def\labelenumii{\alph{enumii}.}
  \item
    \begin{quote}
    This pathway is somewhat likely to increase GPC risk. AI may increase great power involvement and escalation pressure while weakening de-escalation capacity, although proxy deniability, historical success in crisis management, and AI-enabled crisis-management technology could limit the effect.
    \end{quote}
  \end{enumerate}
\item
  \begin{quote}
  \textbf{Strategic environment deterioration.} Patron realignment is already occurring in the Sahel and South Caucasus, disinformation is degrading information environments, and UCDP recorded 61 active state-based conflicts in 2024, the highest since the start of the dataset, 1946\footnote{Rustad, Siri Aas (2025) Conflict Trends: A Global Overview, 1946--2024. \emph{PRIO Paper}. Oslo: PRIO.}.
  \end{quote}

  \begin{enumerate}
  \def\labelenumii{\alph{enumii}.}
  \item
    \begin{quote}
    NGPC is likely to become increasingly destabilizing to great-power relations and strategic environments, particularly through degraded information environments and patron realignment, although the magnitude of destabilisation depends on whether verification technologies and stabilizing institutions keep pace.
    \end{quote}
  \end{enumerate}
\item
  \begin{quote}
  \textbf{AI capability testing and arms race acceleration.} A China-Russia-Iran-North Korea knowledge exchange network is already operational, battlefield-tested AI capabilities are already diffusing, and arms race dynamics are already observable.
  \end{quote}

  \begin{enumerate}
  \def\labelenumii{\alph{enumii}.}
  \item
    \begin{quote}
    NGPC is likely to accelerate military AI competition by demonstrating and diffusing battlefield capabilities, although whether this ultimately increases GPC risk depends partly on whether emerging technologies favor offensive or defensive capabilities.
    \end{quote}
  \end{enumerate}
\end{enumerate}

\emph{Across all three pathways, an increase in the rate of NGPC is likely to increase the probability of great power conflict as AI capabilities advance.}

\subsection*{H2: NGPC Increases the Expected Harm from Catastrophic Terrorism}\label{h2-ngpc-increases-the-expected-harm-from-catastrophic-terrorism}

We use a standard three parameter risk model:

Expected Harm = Attempt Rate $\times$ Success Rate $\times$ Severity.

We investigate three pathways:

\begin{enumerate}
\def\labelenumi{\arabic{enumi}.}
\item
  \begin{quote}
  \textbf{Capability transfer from conflict zones.} Conflict zone pathways add something open-source AI channels cannot-\/- tacit operational knowledge: how to deploy capabilities under real-world conditions, evade countermeasures, and integrate technologies into existing tactics. For example, cartel drone attacks in Mexico rose from 5 in 2020 to 107 in 2021, 233 in 2022, and 260 in the first half of 2023 alone.\footnote{Henry Ziemer, ``Illicit Innovation: Latin America Is Not Prepared to Fight Criminal Drones,'' Center for Strategic and International Studies (CSIS), June 11, 2025, \href{https://www.csis.org/analysis/illicit-innovation-latin-america-not-prepared-fight-criminal-drones}{{https://www.csis.org/analysis/illicit-innovation-latin-america-not-prepared-fight-criminal-drones}}} Cartel-linked operatives have separately been reported to have infiltrated Ukraine's International Legion for drone training, though the attack growth predates the reporting.\footnote{Linus Höller, ``Drug Cartel Operatives Snuck into Ukraine for Drone Training: Report,'' \emph{Defense News}, July 30, 2025, reporting that Ukrainian counterintelligence was investigating alleged infiltration of Ukraine's International Legion by Latin American cartel-linked operatives seeking FPV drone training, based on reporting first published by \emph{Intelligence Online}, \href{https://www.defensenews.com/global/the-americas/2025/07/30/drug-cartel-operatives-snuck-into-ukraine-for-drone-training-report/}{{https://www.defensenews.com/global/the-americas/2025/07/30/drug-cartel-operatives-snuck-into-ukraine-for-drone-training-report/}}, Stephen Honan, ``Drug Cartels Are Adopting Cutting-Edge Drone Technology. Here's How the US Must Adapt,'' \emph{Atlantic Council}, September 29, 2025, \href{https://www.atlanticcouncil.org/blogs/new-atlanticist/drug-cartels-are-adopting-cutting-edge-drone-technology-heres-how-the-us-must-adapt/?utm_source=chatgpt.com}{{https://www.atlanticcouncil.org/blogs/new-atlanticist/drug-cartels-are-adopting-cutting-edge-drone-technology-heres-how-the-us-must-adapt/}}.} What has transferred so far is human skill, not AI knowledge.
  \end{quote}

  \begin{enumerate}
  \def\labelenumii{\alph{enumii}.}
  \item
    \begin{quote}
    Capability transfer is unlikely to increase attempt rate, likely or probable to increase the success rate of catastrophic terrorist attempts in the near term (and highly likely as AI advances), while an increase in severity is a realistic possibility. We are moderately confident in this pathway, since existing transfer mechanisms are well-documented, but the transfer of advanced AI capabilities is diffuse and difficult to predict.
    \end{quote}
  \end{enumerate}
\item
  \begin{quote}
  \textbf{AI-assisted attack planning.} AI combines the adaptability of human guidance with the operational security of working alone, expanding the pool of capable actors (attempt rate) while improving plan quality (success rate). Severity remains limited because AI can help with planning, but major barriers to carrying out an attack remain.
  \end{quote}

  \begin{enumerate}
  \def\labelenumii{\alph{enumii}.}
  \item
    \begin{quote}
    Capable agentic AI and AI-assisted planning are likely or probable to increase both attempt and success rates of attacks by lowering knowledge and execution barriers, while an increase in severity is a realistic possibility, limited by physical-world execution barriers. We are moderately confident in this pathway, since documented AI-assisted plots provide an evidence base, but projections about agentic systems remain forward-looking.
    \end{quote}
  \end{enumerate}
\item
  \begin{quote}
  \textbf{AI-enabled coordination.} AI could compress the decades of sustained training, materiel, and funding relationships required for proxy networks. This pathway has the most uncertainty, since the mechanism is plausible, but the evidence base is thin.
  \end{quote}

  \begin{enumerate}
  \def\labelenumii{\alph{enumii}.}
  \item
    \begin{quote}
    AI-enabled coordination poses a realistic possibility of increasing the rate of attempts, success, and severity in the future. However, our confidence is the lowest in this pathway, since there haven't been historical precedents of AI enabling inter-group coordination that was previously infeasible without AI.
    \end{quote}
  \end{enumerate}
\end{enumerate}

\emph{We conclude that an increased incidence of NGPC is likely or probable to increase the expected harm from catastrophic terrorism, driven primarily by the first two pathways.} Substantial uncertainty remains: the actual effect could be lower than we assess, or significantly higher, particularly for biological catastrophic terrorism.

\subsection*{H3: NGPC Increases the Expected Harm from Loss of Control over Advanced AI}\label{h3-ngpc-increases-the-expected-harm-from-loss-of-control-over-advanced-ai}

This sub-hypothesis is the most speculative of the three. We do not attempt to parameterize the risk. Our main hypothesis is that NGPC increases the expected harm from loss of control by increasing the probability that misaligned or insufficiently controllable AI systems are: (i) deployed or (ii) given risky affordances under (iii) conditions where human detection and mitigation are difficult or costly.

\begin{enumerate}
\def\labelenumi{\arabic{enumi}.}
\item
  \begin{quote}
  \textbf{Deployment pressure.} NGPC may increase the likelihood that misaligned AI systems are deployed by increasing time pressure, resource scarcity, and other incentives that favour hasty capability deployment over safety and alignment. If alignment and safety measures impose meaningful costs in capability, rollout speed, and autonomy, NGPCs may shift deployment toward less aligned systems.
  \end{quote}

  \begin{enumerate}
  \def\labelenumii{\alph{enumii}.}
  \item
    \begin{quote}
    This pathway is plausible but highly conditional. Its risk depends on how steep the safety--performance trade-offs are for future frontier models, and whether costs in speed and capability during NGPCs heavily outweighs actors' preference for reliable and predictable military systems.
    \end{quote}
  \end{enumerate}
\item
  \begin{quote}
  \textbf{Increased affordances.} In fragile states experiencing prolonged conflict, institutional capacity may degrade and incentivize the use of AI systems to substitute for humans in intelligence, coordination, and decision-making roles. Giving AI systems these affordances may increase the likelihood that a misaligned system has increased information, power, and autonomy to carry out hidden objectives.
  \end{quote}

  \begin{enumerate}
  \def\labelenumii{\alph{enumii}.}
  \item
    \begin{quote}
    This pathway is plausible but context-dependent, with the strongest case in prolonged conflicts where human administrative or military capacity is constrained. The effect could be substantially smaller, or even reversed, if AI augments rather than replaces human decision-making and effective human oversight and intervention remain in place.
    \end{quote}
  \end{enumerate}
\item
  \begin{quote}
  \textbf{Detection and mitigation difficulty.} NGPC may make emerging loss of control trajectories harder to identify and contain through two mechanisms. First, NGPC might push decision-makers to accelerate early, path-dependent deployment actions, under the assumption that early deployment may bring about a decisive military advantage. Second, interactions among heterogeneous AI and human agents can make failures harder to predict. These difficulties emerge in three problems: observability, attribution, and intervention.
  \end{quote}

  \begin{enumerate}
  \def\labelenumii{\alph{enumii}.}
  \item
    \begin{quote}
    Mechanisms in this pathway are the most speculative, and its risk relies on the perception of military AI leading to a decisive military advantage. Additionally, AI-enabled monitoring and OSINT may improve observability. The overall pathway is therefore plausible but uncertain and likely to vary across NGPCs.
    \end{quote}
  \end{enumerate}
\end{enumerate}

\emph{We conclude that an increased incidence of NGPC could potentially increase the expected harm and likelihood of loss of control events. However this pathway is the most speculative.}

Importantly, all three loss-of-control pathways are conditional on non-great-power actors gaining access to highly capable AI systems that pose meaningful loss-of-control risk. That access could come through AI arms racing and capability testing, or as a natural consequence of rising AI capabilities and open-weight model releases, though its likelihood and timing remain uncertain.

\textbf{We find the null poorly supported for H1 and H2, and identify H3 as a priority for further work rather than a settled finding.} We conclude that further investigation -- at a minimum, into the cause area's tractability -- is warranted.

We bracket a fourth sub-hypothesis -- that NGPC increases the risk and severity of direct harm from conflict -- as a background assumption. The relevant historical baselines for direct harm from non-great-power conflict are well-established\footnote{Bethany Lacina and Nils Petter Gleditsch, ``Monitoring Trends in Global Combat: A New Dataset of Battle Deaths,'' \emph{European Journal of Population} 21, nos. 2--3 (2005): 145--166; Uppsala Conflict Data Program, ``UCDP Definitions,'' Uppsala University, accessed August 12, 2026.}\footnote{J. David Singer and Melvin Small, \emph{Correlates of War Project: International and Civil War Data, 1816--1992} (Inter-university Consortium for Political and Social Research, January 12, 2006), \href{https://doi.org/10.3886/ICPSR09905.v1}{{https://doi.org/10.3886/ICPSR09905.v1}}.}; AI's contribution to direct harm, while important, does not require the same kind of novel analysis as H1-H3.

\setcounter{section}{0}
\section{Methodology}\label{methodology}

Our approach proceeds in three steps. First, for each sub-hypothesis, we construct a causal model that connects NGPC to the relevant risk outcome through intermediate variables. Second, where evidence permits, we decompose those causal models into parameterized risk models: simple, multiplicative decompositions of the form used in semi-quantitative risk assessment. Third, we perform a first-cut qualitative assessment of each model's parameters, drawing on literature reviews, historical case studies, and abductive inference about how increasingly advanced AI could alter the relevant dynamics. We use the UK intelligence probability yardstick to anchor our qualitative probability assessments in numerical bounds\footnote{UK Government, ``Explaining Uncertainty in UK Intelligence Assessment,'' \emph{GOV.UK}, \href{https://www.gov.uk/government/publications/explaining-uncertainty-in-uk-intelligence-assessment/explaining-uncertainty-in-uk-intelligence-assessment}{{https://www.gov.uk/government/publications/explaining-uncertainty-in-uk-intelligence-assessment/explaining-uncertainty-in-uk-intelligence-assessment}}.}.

We apply this methodology at two levels of rigor, depending on the sub-hypothesis.

For great power conflict escalation and catastrophic terrorism, we construct explicit risk decompositions. Each decomposition breaks the relevant outcome into a product of parameters -- for example, attempt rate, success rate, and severity -- and we qualitatively assess the effect of NGPC on these parameters. We assign qualitative confidence levels to each estimate. These models are not intended to be precise estimates of the risk; they aim to identify the parameters that matter most and to provide a rough sense of the magnitude of the risk.

For loss of control, the mechanisms connecting NGPC to increased risk are currently too speculative to support even rough parameterization without producing false precision. Instead, we identify candidate causal pathways and assess each pathway along three dimensions: \emph{plausibility} (is each link in the causal chain supported by evidence or analogy?), \emph{novelty} (does NGPC contribute something beyond what GPC already contributes to this risk?), and \emph{severity} (if this pathway materializes, how consequential is it?).

Across all three sub-hypotheses, we attempt to identify shared intermediate variables. Several variables, such as the diffusion of AI capabilities to nonstate actors, the erosion of governance norms, and the decay of institutional capacity in conflict-affected states, appear in multiple causal pathways. Identifying these shared nodes is one of the main contributions of this analysis. A variable that appears across multiple risk pathways is, \emph{ceteris paribus}, a higher-priority target for investigation and intervention.

\textbf{Two caveats on interpretation.} First, our probability assessments express confidence about the direction of a parameter change and not its size. When we judge that an increase in success rate is ``highly likely,'' we are making a claim about whether the parameter moves, not about how far it moves. The yardstick was designed for statements about discrete events, and applying it to changes in continuous parameters inherits that limitation. Readers should not take our assessments as implying that the underlying effects are large.

Second, every pathway we assess is conditional on non-great-power and non-state actors gaining access to AI systems capable enough to matter for the mechanisms we describe. That access may come through the arms-race and capability-testing dynamics discussed in Section 2, or simply as a consequence of rising frontier capability and open-weight release, but its timing is uncertain and we do not attempt to forecast it. Where we assess a pathway as likely, that judgment is conditional on the diffusion premise holding. If diffusion is slower or more limited than we assume, the effects we describe are correspondingly delayed or attenuated.

\section{Great Power Conflict Escalation}\label{great-power-conflict-escalation}

\subsection{Introduction}\label{introduction-1}

The claim that NGPC increases the likelihood of great power conflict is, in one sense, uncontroversial: the most destructive great power conflicts of the twentieth century were preceded by or intertwined with smaller conflicts that drew in major powers.\footnote{On the general pattern by which localized disputes escalate into great-power war, see Paul D. Senese and John A. Vasquez, \emph{The Steps to War: An Empirical Study} (Princeton, NJ: Princeton University Press, 2008). On the First World War as a regional conflict that expanded into great-power war, see John A. Vasquez, \emph{Contagion and War: Lessons from the First World War} (Cambridge: Cambridge University Press, 2018); and Christopher Clark, \emph{The Sleepwalkers: How Europe Went to War in 1914} (London: Allen Lane, 2012). On the comparable dynamic in the Asian theater of the Second World War, see John A. Vasquez and Douglas M. Gibler, ``The Steps to War in Asia, 1931--45,'' \emph{Security Studies} 10, no. 3 (2001): 1--45. For an earlier quantitative treatment of war contagion among great powers, see Jack S. Levy, ``The Contagion of Great Power War Behavior, 1495--1975,'' \emph{American Journal of Political Science} 26, no. 3 (1982): 562--584.} What is less clear is the magnitude of this risk relative to other, more direct ways great power conflict could erupt.

In each of the three main pathways we identify, advanced AI is what distinguishes the contemporary risk from its twentieth-century analogue --- not by opening a new route from NGPC to GPC, but by raising escalatory pressures while degrading the mechanisms that have historically contained them. First, NGPC could escalate to GPC directly, through proxy conflict dynamics in which great powers are drawn onto opposing sides of a regional conflict --- a dynamic AI sharpens as autonomous systems and compressed decision timelines erode the deniability and principal--agent slack that once let patrons restrain their proxies. Second, NGPC could increase GPC risk indirectly, by shaping the strategic environment in ways that raise tensions, erode trust, or create commitment problems between great powers --- with AI acting chiefly on the information environment, as disinformation and synthetic media corrode the epistemic basis for crisis communication. Third, NGPC could serve as a testing ground for AI-enabled military capabilities, accelerating arms race dynamics --- the one pathway where AI is constitutive rather than contributory.

\subsection{Risk model}\label{risk-model}

We model the risk of NGPC escalating to great power conflict through two decompositions. The first captures direct escalation: a specific NGPC draws in great powers on opposing sides and escalates through proxy dynamics or alliance entrapment. The second captures indirect escalation: the cumulative effect of NGPCs destabilizes the strategic environment in ways that raise the baseline probability of great power confrontation. Together, the two models cover three pathways: proxy conflict escalation, strategic environment deterioration, and conflict-zone testing of AI-enabled military capabilities.

Model A and Model B capture different causal logics. Model A is event-centric: it asks whether a specific NGPC escalates step-by-step to GPC. Model B is system-centric: it asks whether the cumulative effect of NGPCs raises background GPC risk.

\textbf{Model A: Direct Escalation}

\begin{equation}
\label{eq:modelA}
P(\mathrm{GPC}\mid \mathrm{NGPC}) = R_{\mathrm{NGPC}} \times R_{\mathrm{inv}} \times R_{\mathrm{esc}} \times (1 - R_{\mathrm{de}})
\end{equation}

This model treats direct escalation as a sequential, causal process. Once an NGPC occurs, we examine the probability that great powers may become involved on opposing sides. If great power involvement in NGPCs creates escalation pressures, then those pressures are either successfully de-escalated or they are not. Each parameter is a conditional probability at one stage of this chain. The model applies to both interstate NGPCs and civil conflicts that attract external intervention, and can be evaluated for specific contemporary dyads.

The academic literature on conflict dynamics suggests that these variables should be evaluated based on individual dyads or subsets of countries, rather than as an aggregate predictor of war. The enduring rivalry literature shows that roughly 5\% of dyads account for a disproportionate share of international conflict and wars\footnote{Paul F. Diehl and Gary Goertz, \emph{War and Peace in International Rivalry} (Ann Arbor: University of Michigan Press, 2000), 60--61.}, which suggests that NGPC Rate concentrates in identifiable dyadic relationships rather than arising randomly. Conflict diffusion research further shows that alliance networks and borders function as transmission vectors for conflict spread.\footnote{Randolph M. Siverson and Harvey Starr, \emph{The Diffusion of War: A Study of Opportunity and Willingness} (Ann Arbor: University of Michigan Press, 1991).}

\textbf{Model B: Strategic Deterioration}

\begin{equation}
\label{eq:modelB}
P(\mathrm{GPC}\mid \mathrm{NGPC}) = E_{\mathrm{inst}} \times R_{\mathrm{crisis}} \times R_{\mathrm{c\rightarrow g}}
\end{equation}

This model captures the indirect pathway. NGPC does not escalate to GPC through a direct causal chain, but instead shifts background conditions that raise the probability of great power confrontation through other channels.\footnote{We considered giving the testing-ground pathway its own model. We opted against this because the testing-ground and strategic environment pathways share the same terminal node: both operate by shifting background conditions that feed into existing GPC crisis dynamics. The sub-channel structure within Model B lets us distinguish political from military-technological instability without the structural overhead of a third model.} The NGPC Instability Effect is a multiplier on existing GPC risk: a value of 1.0 means NGPC has no effect; values above 1.0 mean the NGPC environment is net destabilizing for great power relations. The Baseline GP Crisis Rate and Crisis-to-GPC Escalation Rate can be calibrated from existing quantitative work on great power crises.\footnote{Michael Brecher and Jonathan Wilkenfeld, \emph{A Study of Crisis} (Ann Arbor: University of Michigan Press, 1997); Paul K. Huth, \emph{Extended Deterrence and the Prevention of War} (New Haven, CT: Yale University Press, 1988).} In our analysis, we isolate and assess the magnitude and direction of the NGPC Instability Effect alone, taking the other two parameters as given.

The NGPC Instability Effect operates through two mechanisms.

The first mechanism, subchannel A, covers political-strategic instability: trust erosion between great powers (e.g., mutual suspicion over proxy involvement), commitment problems (demonstrated willingness or unwillingness to honor security guarantees), alliance stress and patron realignment,\footnote{Glenn H. Snyder, ``The Security Dilemma in Alliance Politics,'' \emph{World Politics} 36, no. 4 (1984): 461--495.} norm degradation (erosion of sovereignty norms, the laws of war, or the nuclear taboo), and domestic political effects (rally-round-the-flag dynamics, threat inflation, audience cost manipulation).\footnote{See Jessica L. Weeks, ``Autocratic Audience Costs: Regime Type and Signaling Resolve,'' \emph{International Organization} 62, no. 1 (2008): 35--64.}

The second mechanism, subchannel B, covers military-technological instability: the use of NGPCs as testing grounds for AI-enabled weapons systems, which could trigger arms race acceleration through demonstrations of capabilities, or through demonstrations of shifts in the offense-defense balance. The Ukraine conflict illustrates this sub-channel. Battlefield lessons, drone manufacturing cooperation, and component supply chains have produced a coordinated military-technological community among China, Russia, Iran, and North Korea that did not exist before the conflict.\footnote{Bonny Lin et al., ``CRINK Security Ties: Growing Cooperation, Anchored by China and Russia,'' Center for Strategic and International Studies, September 30, 2025, \href{https://www.csis.org/analysis/crink-security-ties-growing-cooperation-anchored-china-and-russia}{{https://www.csis.org/analysis/crink-security-ties-growing-cooperation-anchored-china-and-russia}}.} While the Ukraine war arguably demonstrates defense-dominant technologies, the offense-defense balance could shift in the other direction with future, AI-enabled battlefield technologies.

This distinction matters. A world with many small NGPCs that individually carry low Model A risk could still carry high Model B risk through ratcheting instability effects. Conversely, a single NGPC with high Model A risk (e.g., the South China Sea) might have low Model B risk if it occurs independently. The policy implications also differ: Model A risk is reduced by managing specific conflicts; Model B risk may require systemic interventions, such as arms control, alliance management, and technology governance.

Several variables appear across both models. These are shared nodes in the causal architecture and are, \emph{ceteris paribus}, higher-priority targets for investigation and intervention, because shifting them affects multiple risk pathways at once. See the Appendix for further investigation.

\subsection{Investigation}\label{investigation}

We assess three pathways through which NGPC, combined with increasingly advanced AI, could increase the probability of great power conflict.

\subsubsection{Proxy escalation and alliance entrapment}\label{proxy-escalation-and-alliance-entrapment}

The most direct pathway from NGPC to GPC runs through proxy conflict dynamics. A non-great-power conflict attracts great power involvement on opposing sides; that involvement creates escalation pressures through inadvertent military contact, commitment traps, or alliance entrapment; and those pressures either produce direct great power confrontation or are successfully managed.\footnote{Thomas J. Christensen and Jack Snyder, ``Chain Gangs and Passed Bucks: Predicting Alliance Patterns in Multipolarity,'' \emph{International Organization} 44, no. 2 (1990): 137--168.}

The historical record supports this pathway. During the Cold War, proxy conflicts repeatedly threatened to escalate into superpower confrontation. The Korean War began as a civil conflict and rapidly drew in Chinese and American forces; the Truman-MacArthur confrontation over expanding the war into China illustrates the internal pressure for further escalation.\footnote{Chen Jian, \emph{China's Road to the Korean War: The Making of the Sino-American Confrontation} (New York: Columbia University Press, 1994), 217.} The 1973 Yom Kippur War brought the United States and Soviet Union to DEFCON 3 during a period of supposed detente.\footnote{Barry M. Blechman and Douglas M. Hart, ``The Political Utility of Nuclear Weapons: The 1973 Middle East Crisis,'' \emph{International Security} 7, no. 1 (1982): 132--156.} The Angolan Civil War drew in Cuban, Soviet, and American forces in increasingly dangerous configurations.\footnote{Piero Gleijeses, \emph{Conflicting Missions: Havana, Washington, and Africa, 1959--1976} (Chapel Hill: University of North Carolina Press, 2002).} The mechanism was consistent across cases: rival great powers perceived intervention as weakening their security position and responded by backing favorable parties, producing action-reaction cycles.\footnote{Odd Arne Westad, \emph{The Global Cold War: Third World Interventions and the Making of Our Times} (Cambridge: Cambridge University Press, 2005).}

Several contemporary dyads show this pathway at work. The South China Sea is the most acute. Chinese gray-zone escalation against Philippine vessels near Second Thomas Shoal risks triggering the U.S.-Philippines Mutual Defense Treaty, a direct legal pathway from a gray-zone incident to great power confrontation.\footnote{Council on Foreign Relations, ``Territorial Disputes in the South China Sea,'' Global Conflict Tracker, accessed February 28, 2026, \href{https://www.cfr.org/global-conflict-tracker/conflict/territorial-disputes-south-china-sea}{{https://www.cfr.org/global-conflict-tracker/conflict/territorial-disputes-south-china-sea}}.} The India-Pakistan dyad now, for the first time, features opposing great power patrons (U.S./Quad and China/CPEC), a chain-ganging configuration that did not exist during the Cold War.\footnote{Atlantic Council, ``Experts React: India and Pakistan Have Agreed to a Shaky Cease-Fire. Where Does the Region Go from Here?,'' New Atlanticist, May 2025, \href{https://www.atlanticcouncil.org/blogs/ne\%EE\%80\%80w-atlanticist/experts-react/india-pakist\%EE\%80\%80an-cease-fire-experts/}{{https://www.atlanticcouncil.org/blogs/new-atlanticist/experts-react/india-pakistan-cease-fire-experts/}}.} Sudan's civil war features at least six external state actors providing material support\footnote{Zeinab Mohammed Salih, ``Conflict in Sudan: A Map of Regional and International Actors,'' Wilson Center, December 19, 2024, \href{https://www.wilsoncenter.org/article/conflict-sudan-map-regional-and-international-actors}{{https://www.wilsoncenter.org/article/conflict-sudan-map-regional-and-international-actors}}.}, with a Russian naval base lease at Port Sudan bringing Russian and American naval forces into operational proximity in the Red Sea.\footnote{Denis Staunton, ``Could Sudan's Offer of a Naval Base to Russia Spark an Effort to End the Civil War?,'' Irish Times, December 3, 2025, \href{https://www.irishtimes.com/world/2025/12/03/could-sudans-offer-of-a-naval-base-to-russia-spark-an-effort-to-end-the-civil-war/}{{https://www.irishtimes.com/world/2025/12/03/could-sudans-offer-of-a-naval-base-to-russia-spark-an-effort-to-end-the-civil-war/}}.}

One caveat frames everything that follows. While the contemporary dyads above are still unfolding, the Cold War, the only closed record we have, generated dozens of proxy conflicts and a base rate of proxy-to-GPC escalation of zero. In none of them did escalation pressure actually produce direct superpower war, and the closest call, Korea, ended in a frozen armistice rather than great-power conflict. We take this null result seriously and address it in full below. Here it fixes the question the rest of this section must answer: not whether proxy conflicts generate escalation pressure, which the record shows they do, but whether the mechanisms that contained that pressure during the Cold War will survive AI-enabled conflict.

There are two important counterarguments to proxy escalation: one based on proxy relationships as de-escalation tools, and one based on competing interpretations of historical precedent.

Proxy relationships can function as de-escalation tools. States use non-state proxies precisely to pursue limited objectives while maintaining plausible deniability and avoiding direct confrontation.\footnote{Sam Mullins, ``The Role of Non-State Actors as Proxies in Irregular Warfare and Malign State Influence,'' Irregular Warfare Center, December 9, 2024, \href{https://irregularwarfarecenter.org/publications/research-studies/the-role-of-non-state-actors-as-proxies-in-irregular-warfare-and-malign-state-influence/}{{https://irregularwarfarecenter.org/publications/research-studies/the-role-of-non-state-actors-as-proxies-in-irregular-warfare-and-malign-state-influence/}}.} Russia's use of proxies in Donbas from 2014 to 2022 is instructive: proxy warfare enabled pursuit of strategic objectives below the threshold of direct great power conflict. It was the abandonment of the proxy model in February 2022, not its use, that produced escalation between Russia and NATO forces supporting Ukraine. This suggests GP Involvement Rate and Escalation Rate may be partially inversely correlated: the deeper a great power's involvement, the stronger the deniability mechanisms it builds into the relationship, which in turn reduce escalation risk.

We now return to the caveat flagged above. The Cold War is this paper's primary evidence base for the proxy escalation pathway, but it also provides a strong counterargument: despite dozens of proxy conflicts, none escalated to direct U.S.-Soviet war. If the pathway were as dangerous as the preceding analysis suggests, why did escalation never occur?

The answer lies in Model A's decomposition. A null outcome does not require that escalation pressures were low, only that the De-escalation Rate was high enough to offset them. Our research suggests the other parameters were far from negligible. For example, between 1977 and 1984 U.S. early-warning systems averaged nearly three moderately serious false alarms a week, 1,152 in total, each requiring assessment before it could be ruled out.\footnote{Center for Defense Information, ``Accidental Nuclear War: A Rising Risk?,'' \emph{The Defense Monitor} 15, no. 7 (1986).} Huth's analysis of 58 cases in which great powers committed to defending allies found that those commitments failed to prevent adversary escalation 41.4\% of the time.\footnote{Paul K. Huth, \emph{Extended Deterrence and the Prevention of War} (New Haven, CT: Yale University Press, 1988), 26.} Sagan documented how the rushed alerting of Minuteman missiles at Malmstrom Air Force Base compromised safeguards against unauthorized launch: with only one launch control center operational and no coded locks on the missiles, a single crew held the physical capability, though not the authority, to launch them.\footnote{Scott D. Sagan, \emph{The Limits of Safety: Organizations, Accidents, and Nuclear Weapons} (Princeton, NJ: Princeton University Press, 1993), 81--85.} Proxy wars, fought ostensibly to avoid direct confrontation, nonetheless generated escalation risks of their own: great powers backing opposing sides became entangled through commitment traps and alliance entrapment, and inadvertent military contact between their forces and advisors could raise the danger of a confrontation neither side sought.\footnote{Thomas J. Christensen and Jack Snyder, ``Chain Gangs and Passed Bucks: Predicting Alliance Patterns in Multipolarity,'' \emph{International Organization} 44, no. 2 (1990): 137--168; Odd Arne Westad, \emph{The Global Cold War: Third World Interventions and the Making of Our Times} (Cambridge: Cambridge University Press, 2005).}

As demonstrated above, the Cold War base rate of zero reflects not low risk but successful, and sometimes barely successful, management of continuously high risk. That management rested on bilateral communication channels developed over decades of crises, decision-making timelines measured in hours rather than seconds, principal-agent slack between patrons and proxies that created natural firebreaks, and mutual recognition of nuclear redlines. It is not clear that these same de-escalatory mechanisms will be as effective, or even present, with AI-enabled conflict.\footnote{On the concern that machine-speed decision-making compresses the timelines and human judgment on which crisis de-escalation has historically depended, see Michael C. Horowitz, ``When Speed Kills: Lethal Autonomous Weapon Systems, Deterrence and Stability,'' \emph{Journal of Strategic Studies} 42, no. 6 (2019): 764--788; James Johnson, ``Inadvertent Escalation in the Age of Intelligence Machines: A New Model for Nuclear Risk in the Digital Age,'' \emph{European Journal of International Security} 7, no. 3 (2022): 337--359; and Jürgen Altmann and Frank Sauer, ``Autonomous Weapon Systems and Strategic Stability,'' \emph{Survival} 59, no. 5 (2017): 117--142.}

AI introduces several pressures here. Autonomous systems deployed in proxy conflicts could produce inadvertent escalation through misidentification of targets or unauthorized engagement with great power assets. AI-enabled surveillance and intelligence sharing between patrons and proxies could deepen great power involvement. AI-generated disinformation could degrade the information environment needed for crisis communication and de-escalation.\footnote{Stephan Lewandowsky et al., ``Misinformation, Disinformation, and Violent Conflict: From Iraq and the 'War on Terror' to Future Threats to Peace,'' \emph{American Psychologist} 68, no. 7 (2013): 487--501.} At the same time, AI could improve early warning and crisis management tools, partially offsetting some of these risks.

\emph{Assessment: This pathway primarily affects GP Involvement Rate and Escalation Rate in Model A. The direction of the effect is upward, though its magnitude is uncertain and possibly small. In the near term, GP Involvement Rate is somewhat likely to increase: great powers are already positioning on opposing sides of multiple NGPCs, and AI-enabled proxy capabilities make involvement cheaper and more deniable. As AI-enabled proxies become more capable, involvement is likely to increase further, though the same deniability may partially suppress escalation, since deeper, better-concealed involvement tends to build in the restraint mechanisms discussed above. Escalation Rate is somewhat likely to increase in the near term, driven by autonomous system deployment in proxy conflict zones and compressed decision-making timelines, becoming likely as AI systems produce escalation dynamics that outpace diplomatic response. De-escalation capacity is somewhat likely to deteriorate in both timeframes, as AI-generated disinformation degrades the information environment needed for crisis management; AI-enabled communication and early-warning tools could offset or even reverse this shift if deployed in time. On balance, we assess that this pathway is somewhat likely to increase the probability of great power conflict, but the counterarguments raised above remain unresolved: proxy deniability may brake escalation, the Cold War record shows de-escalation succeeding under high pressure, and AI may aid crisis management as much as it erodes it. Taken together, these mean the expected size of the increase may be small. We are more confident in the direction of the effect than in its magnitude.}

\subsubsection{Strategic environment deterioration}\label{strategic-environment-deterioration}

The second pathway operates indirectly. NGPCs shape the strategic environment in ways that raise tensions, erode trust, and create commitment problems between great powers, increasing the baseline probability of great power crises through channels that may be geographically and temporally distant from the conflicts themselves. This pathway affects Model B's NGPC Instability Effect through sub-channel A. We assess the direction of the effect on that multiplier rather than its magnitude; in each mechanism below, advanced AI enters not by creating a new source of instability but by lowering the cost, and raising the speed and scale, of the practices that degrade the strategic environment.

Several mechanisms connect NGPC to instability.

\textbf{First, NGPCs generate patron realignment dynamics that reshape alliance structures.} The Sahel is the clearest contemporary example. Three military juntas expelled French forces and pivoted to Russian paramilitary partnerships, restructuring the region's security architecture.\footnote{Federica Saini Fasanotti, ``Juntas and Moscow Reshaping Sahel Alliances,'' GIS Reports (Geopolitical Intelligence Services), April 8, 2026, \href{https://www.gisreportsonline.com/r/reshaping-sahel-alliances/}{{https://www.gisreportsonline.com/r/reshaping-sahel-alliances/}}.} Armenia's pivot from Russia toward Western security partnerships after Azerbaijan's 2023 military operation shows how great power distraction, in this case Russia's Ukraine commitment, enables NGPC escalation, which in turn triggers realignment.\footnote{Walter Landgraf and Nareg Seferian, ``A 'Frozen Conflict' Boils Over: Nagorno-Karabakh in 2023 and Future Implications,'' Foreign Policy Research Institute, January 18, 2024, \href{https://www.fpri.org/article/2024/01/a-frozen-conflict-boils-over-nagorno-karabakh-in-2023-and-future-implications/}{{https://www.fpri.org/article/2024/01/a-frozen-conflict-boils-over-nagorno-karabakh-in-2023-and-future-implications/}}.} These cascades increase system-level instability in two ways. New patrons carry untested commitments, creating windows where adversaries may probe, as Azerbaijan did when Russia's Ukraine distraction signaled weakened commitment to Armenia. Departing patrons, meanwhile, face entrapment-abandonment dilemmas: France's withdrawal from the Sahel removed a deterrent without replacing it, opening space for insurgent escalation and further great power competition.\footnote{Glenn H. Snyder, ``The Security Dilemma in Alliance Politics,'' \emph{World Politics} 36, no. 4 (1984): 461--495.} AI's role here is narrower than in the mechanisms that follow, but still present. The diffusion of cheap drones and low-cost autonomous systems has already begun to decouple military reach from industrial scale, allowing nonstate actors and secondary states to field capabilities once reserved for major powers.\footnote{Kerry Chávez and Ori Swed, ``The Proliferation of Drones to Violent Nonstate Actors,'' \emph{Defence Studies} 21, no. 1 (2021): 1--24.} A rising patron can therefore substitute for a departing one at a fraction of the historical cost in personnel and materiel, making realignment faster and more frequent, while AI-enabled surveillance makes the untested commitments of new patrons easier for adversaries to test and exploit.\footnote{Avi Goldfarb and Jon R. Lindsay, ``Prediction and Judgment: Why Artificial Intelligence Increases the Importance of Humans in War,'' \emph{International Security} 46, no. 3 (2022): 7--50.}

\textbf{Second, NGPCs erode international norms that constrain great power behavior by establishing precedents in domains where governance frameworks are weak or absent.} Of the three mechanisms, this is where AI does the most work. When proxy conflicts serve as testing grounds for autonomous weapons, AI-generated disinformation, and arms transfers to designated terrorist organizations, they normalize practices that existing international law has not yet adequately addressed,\footnote{Diana Panke and Ulrich Petersohn, ``Why International Norms Disappear Sometimes,'' \emph{European Journal of International Relations} 18, no. 4 (2012): 719--742; Ryder McKeown, ``Norm Regress: US Revisionism and the Slow Death of the Torture Norm,'' \emph{International Relations} 23, no. 1 (2009): 5--25.} undermining the institutional foundations of world order.\footnote{Stewart Patrick, ``Rules of Order: Assessing the State of Global Governance,'' Carnegie Endowment for International Peace, September 12, 2023, \href{https://carnegieendowment.org/2023/09/12/rules-of-order-assessing-state-of-global-governance-pub-90517}{{https://carnegieendowment.org/2023/09/12/rules-of-order-assessing-state-of-global-governance-pub-90517}}.} The relevant precedents are increasingly AI-specific: the battlefield use of autonomous weapons, including diffusion of the software required to operate unmanned platforms without human control; the use of AI-generated disinformation to destabilize rival states; and the transfer of AI-enabled military capabilities to nonstate actors, a channel already documented in the movement of armed drones from state arsenals to nonstate groups.\footnote{Chávez and Swed, ``The Proliferation of Drones to Violent Nonstate Actors,'' 1--24.} Each of these practices, once normalized in an NGPC where no governance framework constrains it, degrades the normative limits that have historically constrained great power escalation, and does so faster than treaty-based governance can respond. Because norm erosion and information-environment quality recur across several of our pathways, they are among the shared nodes identified in Section 1 as higher-priority targets for investigation and intervention.

The disinformation dimension illustrates how norm erosion in NGPCs feeds back into great power crisis dynamics. When state-sponsored disinformation becomes normalized as a tool of proxy competition, it degrades the shared epistemic environment that great powers rely on for crisis communication and trust-building. Generative AI sharply raises the stakes: synthetic media and deepfakes lower the cost of producing convincing disinformation toward zero, raise its volume, and make attribution and verification between adversaries substantially harder.\footnote{Daniel L. Byman et al., ``Deepfakes and International Conflict'' (Washington, DC: Brookings Institution, 2023).} Research shows that disinformation propagation through social media increases political polarization and drives domestic terrorism,\footnote{James A. Piazza, ``Fake News: The Effects of Social Media Disinformation on Domestic Terrorism,'' \emph{Dynamics of Asymmetric Conflict} 15, no. 1 (2022): 55--77.} while formal modeling demonstrates that propaganda can create systematic barriers to peace by generating disagreement between populations about the odds of winning.\footnote{Petros G. Sekeris, ``Propaganda and Conflict,'' \emph{Games and Economic Behavior} 153 (2025): 569--585.} These findings concern intra-state and intra-population dynamics; our hypothesis is that the same logic extends to the great power dyad, where NGPC-driven disinformation corrodes the informational preconditions for de-escalation between great powers themselves and can raise crisis initiation rates independently of any specific conflict. That extension is directly supported by work on deepfakes in international conflict, which finds that synthetic media can manufacture provocations, obscure attribution, and compress the time available for crisis decision-making between states.\footnote{Byman et al., ``Deepfakes and International Conflict''; Johnson, ``Inadvertent Escalation in the Age of Intelligence Machines,'' 337--359.} The operational scale is already evident in conflict settings, where 75\% of UN peacekeepers report that disinformation affects their safety, an indication of how far synthetic information has penetrated the active conflict environments in which great power interests intersect.\footnote{Pernilla Rydén et al., ``Tackling Mis- and Disinformation: Seven Insights for UN Peace Operations,'' SIPRI (blog), 2023, \href{https://www.sipri.org/commentary/blog/2023/tackling-mis-and-disinformation-seven-insights-un-peace-operations}{{https://www.sipri.org/commentary/blog/2023/tackling-mis-and-disinformation-seven-insights-un-peace-operations}}.}

\textbf{Third, NGPC outcomes shape great power threat perceptions and domestic politics.} The regime type literature shows that personalist dictatorships are significantly more likely to initiate militarized disputes,\footnote{Jessica L. Weeks, ``Strongmen and Straw Men: Authoritarian Regimes and the Initiation of International Conflict,'' \emph{American Political Science Review} 106, no. 2 (2012): 326--347.} and that autocracies in the developing world are more prone to political instability because oppressed citizenries have no mechanism to voice dissent other than active mobilization.\footnote{Rollin F. Tusalem, ``Democracies, Autocracies, and Political Stability,'' \emph{International Social Science Review} 90, no. 1 (2015).} If NGPCs produce state failure or regime change that increases the number of personalist or unstable autocracies in the system, the baseline expected conflict rate is likely to rise, and the trend is not reassuring: the UCDP recorded 61 active state-based conflicts in 2024, the highest number since data collection began in 1946.\footnote{Shawn Davies et al., ``Organized Violence 1989--2024, and the Challenges of Identifying Civilian Victims,'' \emph{Journal of Peace Research} 62, no. 4 (2025): 1223--1240.} AI cuts ambiguously here. Advanced surveillance, censorship, and predictive-policing tools have become instruments of digital repression that help personalist regimes monitor and suppress dissent, potentially sustaining precisely the regime types most prone to initiating disputes, even as the same tools could, in other institutional settings, strengthen state capacity rather than entrench autocracy.\footnote{Steven Feldstein, \emph{The Rise of Digital Repression: How Technology Is Reshaping Power, Politics, and Resistance} (New York: Oxford University Press, 2021).}

The mechanisms above are not uniformly destabilizing, and several countervailing dynamics could dampen the effect. Norms sometimes prove resilient rather than eroding. The same AI tools that degrade the information environment could also improve it: emerging content-provenance standards, such as the C2PA Content Credentials framework now being adopted for AI-generated media, aim to make manipulated content detectable and to restore attribution, which would raise the cost of disinformation rather than lower it.\footnote{National Security Agency et al., ``Content Credentials: Strengthening Multimedia Integrity in the Generative AI Era,'' Cybersecurity Information Sheet, January 29, 2025, \href{https://media.defense.gov/2025/Jan/29/2003634788/-1/-1/0/CSI-CONTENT-CREDENTIALS.PDF}{{https://media.defense.gov/2025/Jan/29/2003634788/-1/-1/0/CSI-CONTENT-CREDENTIALS.PDF}}.} Patron realignment can consolidate clearer spheres of influence as easily as it can destabilize them. These possibilities do not neutralize the pathway, but they widen the uncertainty around its magnitude.

\emph{Assessment: This pathway affects the NGPC Instability Effect in Model B (sub-channel A). As with the proxy pathway, the direction of the effect is clearer than its magnitude. Political-strategic instability is somewhat likely to increase in the near term: patron realignment is already occurring in the Sahel and South Caucasus, and disinformation is already degrading information environments. In the future this becomes likely, as AI-generated deepfakes and synthetic media complicate attribution and verification between great powers and AI-enabled information warfare operates at a scale and speed that existing institutional response infrastructure appears ill-equipped to match.}\footnote{Byman et al., ``Deepfakes and International Conflict.''} \emph{The magnitude remains uncertain, however, because it depends on whether AI-enabled verification and norm-building keep pace with AI-enabled disruption, and on whether the countervailing dynamics above materialize. We are more confident that NGPC degrades the strategic environment than we are about how large that effect will be.}

\subsubsection{AI capability testing and arms race acceleration}\label{ai-capability-testing-and-arms-race-acceleration}

The third pathway treats NGPCs as innovation hubs. AI-enabled military capabilities are developed there under adaptive pressure, spread to state and non-state actors, and generate demand signals that accelerate arms racing between great powers. This pathway affects Model B's NGPC Instability Effect through sub-channel B. The mechanism is familiar from the arms-race and security-dilemma literature: when one state fields or demonstrates a new capability, rivals feel pressure to match it, and the resulting action-reaction cycles can harden threat perceptions and, under some conditions, raise the probability of war.\footnote{Robert Jervis, ``Cooperation Under the Security Dilemma,'' \emph{World Politics} 30, no. 2 (1978): 167--214; Charles L. Glaser, ``The Causes and Consequences of Arms Races,'' \emph{Annual Review of Political Science} 3 (2000): 251--276.} Our claim has two steps, and they are not equally secure. The first is that deployment in NGPCs speeds the development and diffusion of AI-enabled capabilities. The second is that this acceleration feeds great-power arms racing in ways that raise GPC risk. The evidence for the first is strong; the second, as we argue below, is real but contested in both direction and size.

The conflict in Ukraine is the clearest case of an NGPC serving as a laboratory for AI-enabled warfare. Both sides have used the war to develop AI-assisted targeting, drone autonomy, electronic-warfare countermeasures, and swarm tactics, refined under combat conditions no test range can reproduce.\footnote{Samuel Bendett and David Kirichenko, ``Battlefield Drones and the Accelerating Autonomous Arms Race in Ukraine,'' Center for a New American Security, January 10, 2025, \href{https://www.cnas.org/publications/commen\%EE\%80\%80tary/battlefield-drones-and-the-accelera\%EE\%80\%80ting-autonomous-arms-race-in-ukraine}{{https://www.cnas.org/publications/commentary/battlefield-drones-and-the-accelerating-autonomous-arms-race-in-ukraine}}.} Russia's Lancet loitering munitions carry Nvidia Jetson modules for autonomous target recognition in their terminal phase;\footnote{Institute for Science and International Security, ``Russian Lancet-3 Kamikaze Drone Filled with Foreign Parts,'' December 18, 2023, \href{https://isis-online.org/isis-reports/russian-lancet-3-kamikaze-drone-filled-with-foreign-parts}{{https://isis-online.org/isis-reports/russian-lancet-3-kamikaze-drone-filled-with-foreign-parts}}.} Ukraine's Avengers system uses machine vision to scan drone footage and flags roughly 12,000 Russian targets a week.\footnote{``Ukrainian Troops Detect 12,000 Targets Weekly with Help of AI,'' \emph{Ukrainska Pravda}, September 23, 2024, \href{https://www.pravda.com.ua/eng/news/2024/09/23/7476410/}{{https://www.pravda.com.ua/eng/news/2024/09/23/7476410/}}, quoting Deputy Minister of Defence Kateryna Chernohorenko.} Fully autonomous swarms have not been fielded yet, but the components are maturing quickly, and the war has compressed development cycles from years to months. These capabilities are being proven in combat rather than on test ranges, which collapses the gap between prototype and demonstrated effect and shortens every rival's planning horizon.

The capabilities have not stayed in Ukraine. A military-technological learning community has formed among China, Russia, Iran, and North Korea, whose members trade battlefield lessons, co-develop systems, and coordinate procurement and sanctions evasion.\footnote{Bonny Lin et al., ``CRINK Security Ties: Growing Cooperation, Anchored by China and Russia,'' Center for Strategic and International Studies, September 30, 2025, \href{https://www.csis.org/analysis/crink-secu\%EE\%80\%80rity-ties-growing-cooperation-anchored-c\%EE\%80\%80hina-and-russia}{{https://www.csis.org/analysis/crink-security-ties-growing-cooperation-anchored-china-and-russia}}.} China supplies an estimated 80 percent of the critical electronics in Russian drones;\footnote{David Kirichenko, ``The Booming China-Russia Drone Alliance,'' Center for European Policy Analysis, June 4, 2025, \href{https://cepa.org/article/the-bo\%EE\%80\%80oming-china-russia-drone-alliance/}{{https://cepa.org/article/the-booming-china-russia-drone-alliance/}}.} North Korea has agreed to send up to twelve thousand workers to build Shahed-type drones at Russia's Alabuga Special Economic Zone;\footnote{Rose Adams, ``Alabuga: The Latest Destination for North Korea's Drone Ambitions,'' 38 North, December 17, 2025, \href{https://www.38north.org/2025/12/alabuga-the-latest-destination-for-north-koreas-drone-ambitions/}{{https://www.38north.org/2025/12/alabuga-the-latest-destination-for-north-koreas-drone-ambitions/}}.} and Iranian Shahed designs are tested in Ukraine, with the lessons feeding back into systems later handed to the Houthis and Hezbollah.\footnote{Michael Knights and Alex Almeida, ``What Iran's Drones in Ukraine Mean for the Future of War,'' Washington Institute for Near East Policy, November 10, 2022, \href{https://www.washingtoninstitute.org/poli\%EE\%80\%80cy-analysis/what-irans-drones-ukraine-me\%EE\%80\%80an-future-war}{{https://www.washingtoninstitute.org/policy-analysis/what-irans-drones-ukraine-mean-future-war}}.} The same routes reach non-state actors, moving armed drones out of state arsenals and into insurgent hands.\footnote{Kerry Chávez and Ori Swed, ``The Proliferation of Drones to Violent Nonstate Actors,'' \emph{Defence Studies} 21, no. 1 (2021): 1--24} This capability diffusion is one of the cross-cutting variables identified in Section 1; it feeds this pathway and the proliferation dynamics under H2 at once.

Demonstrated capability in an NGPC creates demand signals for rivals. As drones proved decisive in Ukraine, militaries elsewhere moved money into autonomous systems, counter-drone defenses, and AI-integrated command and control.\footnote{Samuel Bendett and David Kirichenko, ``Battlefield Drones and the Accelerating Autonomous Arms Race in Ukraine,'' Center for a New American Security, January 10, 2025, \href{https://www.cnas.org/publications/commen\%EE\%80\%80tary/battlefield-drones-and-the-accelera\%EE\%80\%80ting-autonomous-arms-race-in-ukraine}{{https://www.cnas.org/publications/commentary/battlefield-drones-and-the-accelerating-autonomous-arms-race-in-ukraine}}.} Whether that spending is destabilizing depends on how AI reshapes the balance of military power. Horowitz argues that AI, as a general-purpose technology rather than a discrete weapon, could tilt the balance toward first movers and toward states able to integrate it at scale.\footnote{Michael C. Horowitz, ``Artificial Intelligence, International Competition, and the Balance of Power,'' \emph{Texas National Security Review} 1, no. 3 (May 2018): 36--57.} A 2026 RAND assessment makes a sharper claim: as AI-enabled uncrewed systems grow cheaper and more capable, the cost-effectiveness of quantity may rise, favoring ``affordable mass'' and the industrial powers that can produce it, though the same report cautions that sophisticated platforms will keep their place.\footnote{Zachary Burdette et al., \emph{How Artificial Intelligence Could Reshape Four Essential Competitions in Future Warfare} (Santa Monica, CA: RAND Corporation, 2026), RR-A4316-1, \href{https://www.rand.org/pubs/re\%EE\%80\%80search_reports/RRA4316-1.html}{{https://www.rand.org/pubs/research\_reports/RRA4316-1.html}}.} If leaders come to believe that cheap autonomous mass is decisive, the pressure to out-produce a rival grows, which is the ordinary engine of an arms race.

Whether that racing actually raises the odds of great-power war is the harder question, and three links connect them. If AI capabilities are read as favoring the attacker, they create first-strike and preventive-war incentives, since each side fears that waiting hands the advantage to the other.\footnote{Robert Jervis, ``Cooperation Under the Security Dilemma,'' \emph{World Politics} 30, no. 2 (1978): 167--214; Ben Garfinkel and Allan Dafoe, ``How Does the Offense-Defense Balance Scale?'' \emph{Journal of Strategic Studies} 42, no. 6 (2019): 736--763.} If they compress the time available for human decision in a crisis, they raise the risk of inadvertent or premature escalation.\footnote{Michael C. Horowitz, ``When Speed Kills: Lethal Autonomous Weapon Systems, Deterrence and Stability,'' \emph{Journal of Strategic Studies} 42, no. 6 (2019): 764--788; Jürgen Altmann and Frank Sauer, ``Autonomous Weapon Systems and Strategic Stability,'' \emph{Survival} 59, no. 5 (2017): 117--142.} Additionally, empirically, arms races are most dangerous inside an existing rivalry, where they correlate with a markedly higher probability of war.\footnote{Toby J. Rider, Michael G. Findley, and Paul F. Diehl, ``Just Part of the Game? Arms Races, Rivalry, and War,'' \emph{Journal of Peace Research} 48, no. 1 (2011): 85--100.} The current AI competition sits squarely inside the U.S.-China rivalry, which is where the historical record is least reassuring.

The pathway is more uncertain than that chain implies, and three objections carry weight. The offense-defense direction is genuinely unsettled: Ukraine arguably points the other way, since cheap drones and saturated sensor coverage have made massed armored assault ruinously expensive, and a defensive advantage is stabilizing rather than destabilizing.\footnote{Ben Garfinkel and Allan Dafoe, ``How Does the Offense-Defense Balance Scale?'' \emph{Journal of Strategic Studies} 42, no. 6 (2019): 736--763.} ``AI arms race'' may also be the wrong label. Scharre argues the competition fails the standard test for one: AI is a diffuse general-purpose technology rather than a countable weapon, military AI spending sits far below the abnormal growth rates that mark real arms races, and the dynamic looks more like an ordinary security dilemma.\footnote{Paul Scharre, ``Debunking the AI Arms Race Theory,'' \emph{Texas National Security Review} 4, no. 3 (Summer 2021): 121--132.} Likewise, arms races do not reliably cause war; the link runs through rivalry, the offense-defense balance, and whether the competition is genuinely reciprocal or just parallel modernization.\footnote{Charles L. Glaser, ``The Causes and Consequences of Arms Races,'' \emph{Annual Review of Political Science} 3 (2000): 251--276; on the rivalry channel specifically, Rider, Findley, and Diehl, ``Just Part of the Game?''} Much of today's investment in autonomous systems may owe little to any particular NGPC demonstration.

\emph{Assessment. This pathway affects the NGPC Instability Effect in Model B through sub-channel B. As with the other pathways, we are surer of the direction than the magnitude, and here the gap is at its widest. That NGPCs speed the development and diffusion of AI-enabled capability is close to certain and already visible: the China-Russia-Iran-North Korea learning community is operational, battlefield-tested systems are spreading through established routes, and defense spending on autonomous systems is rising across several states. Whether that acceleration raises the probability of great-power war is genuinely open. It turns on the offense-defense balance, which current evidence may tilt toward defense; on whether this is a real arms race or ordinary modernization; and on whether racing ends in war or in stable deterrence. We therefore judge that NGPC-driven capability diffusion and arms-race dynamics are likely to intensify, while treating the resulting rise in GPC probability as plausible but uncertain and possibly modest. That is a narrower and better-hedged claim than ``the arms-racing effect is likely to occur,'' and it is the one the evidence supports.}

\subsubsection{Conclusion and summary of analysis}\label{conclusion-and-summary-of-analysis}

Across all three pathways, the direction of the effect points the same way: an increase in the rate and intensity of NGPC is likely to raise the probability of great power conflict as AI capabilities advance. As throughout this section, we are more confident in that direction than in its magnitude, and the three pathways do not carry equal weight.

The proxy escalation pathway creates direct escalation risks in specific high-salience dyads such as the South China Sea, but its expected magnitude may be small. Proxy deniability can brake escalation, the Cold War record shows de-escalation succeeding under sustained pressure, and AI may aid crisis management even as it erodes it. We judge this pathway somewhat likely to increase GPC probability.

The strategic environment pathway raises background tension across the great power system through patron realignment and a degraded information environment. Here the direction is clearer: the instability effect is somewhat likely to operate in the near term and likely as synthetic media matures, though its size depends on whether AI-enabled verification and norm-building keep pace with AI-enabled disruption.

The testing-ground pathway is where our confidence divides most sharply. That NGPCs speed the development and diffusion of AI-enabled capability is close to certain and already visible. Whether that acceleration raises the probability of great-power war is genuinely open, turning on the offense-defense balance, on whether this is a real arms race or ordinary modernization, and on whether racing ends in war or in stable deterrence. We therefore treat the intensification of arms-race dynamics as likely while treating the resulting rise in GPC probability as plausible but uncertain and possibly modest.

The pathways may also interact. A testing-ground dynamic (pathway 3) could increase the capability of proxies (pathway 1), raising the stakes of patron involvement and feeding the strategic instability of pathway 2. They may also share drivers rather than simply reinforce one another. AI capability diffusion appears in pathway 3 here and again in the capability transfer pathway under H2, so treating those two assessments as independent risks counting the same underlying trend twice. We model neither effect. Our independent treatment therefore understates the aggregate where pathways compound and overstates it where they share a common cause, and we do not attempt to determine which dominates. The Appendix identifies the shared drivers so that readers can judge where the double-counting risk lies.

{\footnotesize\setlength{\extrarowheight}{2.5pt}\renewcommand{\arraystretch}{1.2}
\begin{longtable}[]{@{}
  >{\raggedright\arraybackslash}p{(\columnwidth - 12\tabcolsep) * \real{0.1571}}
  >{\raggedright\arraybackslash}p{(\columnwidth - 12\tabcolsep) * \real{0.1426}}
  >{\raggedright\arraybackslash}p{(\columnwidth - 12\tabcolsep) * \real{0.1266}}
  >{\raggedright\arraybackslash}p{(\columnwidth - 12\tabcolsep) * \real{0.1394}}
  >{\raggedright\arraybackslash}p{(\columnwidth - 12\tabcolsep) * \real{0.1298}}
  >{\raggedright\arraybackslash}p{(\columnwidth - 12\tabcolsep) * \real{0.1603}}
  >{\raggedright\arraybackslash}p{(\columnwidth - 12\tabcolsep) * \real{0.1442}}@{}}
\caption{Future effects on the great power conflict risk parameters as AI capabilities advance.}\label{tab:gpc-parameters}\\
\toprule\noalign{}
\begin{minipage}[t]{\linewidth}\raggedright
\textbf{Pathway}
\end{minipage} & \begin{minipage}[t]{\linewidth}\raggedright
\textbf{Primary channel}
\end{minipage} & \begin{minipage}[t]{\linewidth}\raggedright
\textbf{NGPC Rate}
\end{minipage} & \begin{minipage}[t]{\linewidth}\raggedright
\textbf{GP Involvement Rate}
\end{minipage} & \begin{minipage}[t]{\linewidth}\raggedright
\textbf{Escalation Rate}
\end{minipage} & \begin{minipage}[t]{\linewidth}\raggedright
\textbf{De-escalation Capacity}
\end{minipage} & \begin{minipage}[t]{\linewidth}\raggedright
\textbf{NGPC Instability Effect}
\end{minipage} \\\addlinespace
\begin{minipage}[t]{\linewidth}\raggedright
Proxy Escalation
\end{minipage} & \begin{minipage}[t]{\linewidth}\raggedright
Model A
\end{minipage} & \begin{minipage}[t]{\linewidth}\raggedright
N/A
\end{minipage} & \begin{minipage}[t]{\linewidth}\raggedright
Likely
\end{minipage} & \begin{minipage}[t]{\linewidth}\raggedright
Likely
\end{minipage} & \begin{minipage}[t]{\linewidth}\raggedright
Somewhat Likely (deterioration)
\end{minipage} & \begin{minipage}[t]{\linewidth}\raggedright
N/A
\end{minipage} \\\addlinespace
\begin{minipage}[t]{\linewidth}\raggedright
Strategic Environment
\end{minipage} & \begin{minipage}[t]{\linewidth}\raggedright
Model B, sub-channel A
\end{minipage} & \begin{minipage}[t]{\linewidth}\raggedright
Somewhat Likely
\end{minipage} & \begin{minipage}[t]{\linewidth}\raggedright
N/A
\end{minipage} & \begin{minipage}[t]{\linewidth}\raggedright
N/A
\end{minipage} & \begin{minipage}[t]{\linewidth}\raggedright
N/A
\end{minipage} & \begin{minipage}[t]{\linewidth}\raggedright
Likely
\end{minipage} \\\addlinespace
\begin{minipage}[t]{\linewidth}\raggedright
Testing Ground / Arms Race
\end{minipage} & \begin{minipage}[t]{\linewidth}\raggedright
Model B, sub-channel B
\end{minipage} & \begin{minipage}[t]{\linewidth}\raggedright
Somewhat Likely
\end{minipage} & \begin{minipage}[t]{\linewidth}\raggedright
N/A
\end{minipage} & \begin{minipage}[t]{\linewidth}\raggedright
N/A
\end{minipage} & \begin{minipage}[t]{\linewidth}\raggedright
Somewhat Likely (deterioration)
\end{minipage} & \begin{minipage}[t]{\linewidth}\raggedright
Somewhat Likely
\end{minipage} \\\addlinespace
\midrule\noalign{}
\endhead
\bottomrule\noalign{}
\endlastfoot
\end{longtable}
}

\emph{Note: De-escalation capacity deterioration means the parameter moves in the direction that increases risk. Assessments reflect future effects as AI capabilities advance; near-term assessments are generally one step lower (e.g., ``likely'' becomes ``somewhat likely''). Assessments express confidence in the direction of change, not its magnitude.}

\section{Catastrophic Terrorism}\label{catastrophic-terrorism}

\subsection{Introduction}\label{introduction-2}

There is consensus that increasingly advanced AI would increase the risk that nonstate actors are able to commit acts of catastrophic terrorism.\footnote{Jim Mitre and Joel B. Predd, \emph{Artificial General Intelligence's Five Hard National Security Problems} (Santa Monica, CA: RAND Corporation, 2025), \href{https://www.rand.org/pubs/perspectives/PEA3691-4.html}{{https://www.rand.org/pubs/perspectives/PEA3691-4.html}}. The authors identify ``nonexperts empowered to develop weapons of mass destruction'' as one of five hard national security problems posed by AI, noting that AI lowers barriers across chemical, biological, and cyber domains. See also United Nations Office of Counter-Terrorism and United Nations Interregional Crime and Justice Research Institute, \emph{Algorithms and Terrorism: The Malicious Use of Artificial Intelligence for Terrorist Purposes} (New York: United Nations, 2021). In a survey of 27 experts from government, industry, academia, and international organizations, 44 percent assessed malicious use of AI for terrorist purposes as ``very likely'' and 56 percent as ``somewhat likely''; no participants assessed it as unlikely.} Here, we investigate a neglected aspect of that risk: would an increased incidence of NGPC increase the expected harm from catastrophic terrorism?

The intuition is seemingly straightforward: NGPC is where non-state actors operate, and often where states arm them. Unlike GPC, NGPC occurs in under-governed spaces accessible to foreign fighters.\footnote{Martha Crenshaw, ``Transnational Jihadism and Civil Wars,'' \emph{Dædalus} 146, no. 4 (Fall 2017): 59-70, \href{https://www.amacad.org/publication/transnational-jihadism-civil-wars}{{https://www.amacad.org/publication/transnational-jihadism-civil-wars}}. Crenshaw observes that ``sanctuary in failed states or ungoverned spaces is only part of the story. Civil conflicts {[}what we are calling NGPC{]} are attractions for foreign fighters from the region or from distant countries.'' See also Hanna Samir Kassab, \emph{Power Vacuums and Global Politics: Areas of State and Non-state Competition in Multipolarity} (New York: Routledge, 2023), Kassab defines a power vacuum as the fundamental absence of legitimate state authority over a geographic territory, noting that with no state authority governing a geographical region, opportunistic states and organized criminal and terrorist networks may attempt to control that space.} Several concerning dynamics operate in these spaces: operatives travel to war zones to acquire skills, states transfer capabilities to proxies, oversight over what capabilities are developed is limited.

The empirical record supports this intuition. Recent editions of the Global Terrorism Index find that roughly nine in ten terrorist attacks and 98 percent of terrorism deaths occur in countries in conflict, and that attacks there are roughly five to six times deadlier than in peaceful countries.\footnote{Institute for Economics and Peace, \emph{Global Terrorism Index 2024} (Sydney: IEP, 2024), \url{https://www.visionofhumanity.org/maps/global-terrorism-index/}. The GTI draws on Dragonfly's TerrorismTracker database.} The vast majority of these conflict-affected countries are experiencing non-great-power conflicts --- from insurgencies in the Sahel and Myanmar, to proxy wars in Sudan and Yemen, to cartel violence in Mexico and Colombia. The causal channels are well-documented: states experiencing chronic failure are significantly more likely to host and produce transnational terrorist groups,\footnote{James A. Piazza, ``Incubators of Terror: Do Failed and Failing States Promote Transnational Terrorism?'' International Studies Quarterly 52, no. 3 (2008): 469--488{,} {\url{https://doi.org/10.1111/j.1468-2478.2008.00511.x}.} Piazza's time-series, cross-national negative binomial analysis of 197 countries from 1973 to 2003 finds that states plagued by chronic state failures are statistically more likely to host terrorist groups that commit transnational attacks, have their nationals commit transnational attacks, and are more likely to be targeted by transnational terrorists themselves. See also Michael G. Findley and Joseph K. Young, ``Terrorism and Civil War: A Spatial and Temporal Approach to a Conceptual Problem,'' Perspectives on Politics 10, no. 2 (2012): 285-305{,} \href{https://doi.org/10.1017/S1537592712000679}{{https://doi.org/10.1017/S1537592712000679.}}} and the weaker a government's territorial control, the more terrorism it produces.\footnote{Brian Lai, ``'Draining the Swamp': An Empirical Examination of the Production of International Terrorism, 1968--1998,'' Conflict Management and Peace Science 24, no. 4 (2007): 297--310\textsuperscript{,} {\url{https://doi.org/10.1080/07388940701643649}.} Lai finds strong support for the state strength approach: the lower a state's ability to impose costs on terrorist groups within its borders, the greater the amount of terrorism produced within that state.} While the causal direction is debated, recent analysis confirms that state weakness drives terrorist group survival, not solely the reverse.\footnote{Khusrav Gaibulloev, James A. Piazza, and Todd Sandler, ``Do Failed or Weak States Favor Resident Terrorist Groups' Survival?'' Journal of Conflict Resolution 68, no. 5 (2024): 823--848, \href{https://doi.org/10.1177/00220027231183939}{{https://doi.org/10.1177/00220027231183939}}. The authors note that ``endogeneity, stemming from reverse causality and omitted variables, presents a threat to identifying the causal relationship'' and apply an instrumental variables approach to address it, finding that weak states do favor resident terrorist groups' longevity. See also Khusrav Gaibulloev and Todd Sandler, ``Are Resident Terrorist Groups Productive in Weak States?'' \emph{Kyklos} (2025), \href{https://doi.org/10.1111/kykl.12458}{{https://doi.org/10.1111/kykl.12458}}, which confirms that terrorist groups in weak states generate more attacks across all categories.}

Conflict zones also generate terrorism through a foreign fighter pipeline that operates across ideological lines. Hegghammer's analysis found that one in nine (or less) of Western jihadist foreign fighters returned to take part in domestic terror plots, and that such plots were more likely to produce fatalities.\footnote{Thomas Hegghammer, ``Should I Stay or Should I Go? Explaining Variation in Western Jihadists' Choice between Domestic and Foreign Fighting,'' \emph{American Political Science Review} 107, no. 1 (2013): 1--15, \href{https://doi.org/10.1017/S0003055412000615}{{https://doi.org/10.1017/S0003055412000615}}.} The same dynamic appears in far-right terrorism: the Ukraine conflict has provided combat skills and transnational networks to European extremists, at least one of whom was convicted of planning attacks in France using weapons acquired in eastern Ukraine.\footnote{On far-right foreign fighters and Ukraine, see Christian Kaunert, Alex MacKenzie, and Sarah Léonard, ``Far-right Foreign Fighters and Ukraine: A Blind Spot for the European Union?'' \emph{New Journal of European Criminal Law 14, no. 2 (2023):} 247--266, \href{https://doi.org/10.1177/20322844231164089}{{https://doi.org/10.1177/20322844231164089}}; and Kacper Rękawek, \emph{Foreign Fighters in Ukraine} (London: Routledge, 2023). The convicted individual is Grégoire Moutaux, a French far-right extremist who entered Ukraine in June 2015 and came to SBU attention in December 2015. He offered a Ukrainian serviceman €16,000 for five Kalashnikov rifles, two rocket-propelled grenade launchers, 125 kilograms of TNT, and 100 electric detonators; the serviceman alerted the SBU, which allowed the purchase to proceed as a controlled operation. Ukraine's SBU arrested him in May 2016 at the Polish-Ukrainian border; he had planned 15 attacks targeting mosques, synagogues, bridges, and infrastructure during the Euro 2016 football championship. He was convicted in Ukraine in 2018 on terrorism and smuggling charges. See ``Ukraine Convicts Frenchman over Euro 2016 Attack Plot,'' France 24, May 22, 2018, \href{https://www.france24.com/en/20180522-ukraine-terrorism-convicts-frenchman-over-euro-2016-football-france-attack-plot}{{https://www.france24.com/en/20180522-ukraine-terrorism-convicts-frenchman-over-euro-2016-football-france-attack-plot}}; ``French Suspect Arrested for Alleged Attack Plot,'' \emph{Washington Post}, June 6, 2016, \href{https://www.washingtonpost.com/world/french-suspect-arrested-for-alleged-attack-plot-during-soccer-tournament-ukraine-says/2016/06/06/199d7c73-9546-421c-89af-1a1d59989fc1_story.html}{{https://www.washingtonpost.com/world/french-suspect-arrested-for-alleged-attack-plot-during-soccer-tournament-ukraine-says/2016/06/06/199d7c73-9546-421c-89af-1a1d59989fc1\_story.html}}. For broader analysis, see New Lines Institute, ``The Balkan Connection: Foreign Fighters and the Far Right in Ukraine,'' September 2023, \href{https://newlinesinstitute.org/nonstate-actors/the-balkan-connection-foreign-fighters-and-the-far-right-in-ukraine/}{{https://newlinesinstitute.org/nonstate-actors/the-balkan-connection-foreign-fighters-and-the-far-right-in-ukraine/}}.} The cartel drone pipeline (documented below) represents a third variant that is operational rather than ideological. Earlier pipelines followed the same logic: the IRA acquired weapons from Libya, FARC developed extensive military and illicit-financing capabilities over decades of Colombian civil war, and the Tamil Tigers further refined suicide-bombing tactics, including suicide belts and the systematic use of female attackers; similar tactics were subsequently adopted by armed groups elsewhere.\footnote{On the IRA--Libya connection, see Ed Moloney, \emph{A Secret History of the IRA} (New York: W.W. Norton, 2002). On FARC, see James J. Brittain, \emph{Revolutionary Social Change in Colombia: The Origin and Direction of the FARC-EP} (London: Pluto Press, 2010). On the spread of suicide bombing tactics refined by the Tamil Tigers, see Robert A. Pape, \emph{Dying to Win: The Strategic Logic of Suicide Terrorism} (New York: Random House, 2005); Michael C. Horowitz, ``Nonstate Actors and the Diffusion of Innovations: The Case of Suicide Terrorism,'' \emph{International Organization} 64, no. 1 (2010): 33--64.} In each case, conflict zones provide what peacetime training cannot: hands-on, practical knowledge of how to deploy capabilities under real-world conditions, evade countermeasures, and integrate emerging technologies into existing tactics.

What is less clear is whether these dynamics meaningfully change the risk of catastrophic terrorist attacks. Historically, implementation barriers have dominated: groups still need resources, organizational capacity, and the ability to move from planning to execution to perform an attack.\footnote{National Academies of Sciences, Engineering, and Medicine, \emph{Nuclear Terrorism: Assessment of U.S. Strategies to Prevent, Counter, and Respond to Weapons of Mass Destruction} (Washington, DC: National Academies Press, 2024), \href{https://nap.nationalacademies.org/catalog/27215/}{{https://nap.nationalacademies.org/catalog/27215/}}. The report notes that ``in order for a terrorist organization to carry out a nuclear attack with either a nuclear weapon or an improvised nuclear device, they would need the complicity of a state, the failure of the state's controls, or the failure of the state itself'' (Chapter 3, ``New Dynamics in Nuclear Terrorism Pose New Risks'').} Knowledge of how to build a weapon has rarely been the only limiting factor. The desire to commit such an attack is also a barrier: terrorist groups that maintain close ties to local populations are constrained by those relationships, since indiscriminate and catastrophic violence risks alienating the very communities they depend on for recruits, resources, and legitimacy.\footnote{James A. Piazza, ``Is Islamist Terrorism More Dangerous?: An Empirical Study of Group Ideology, Organization, and Goal Structure,'' Terrorism and Political Violence 21, no. 1 (2009): 62--88, \href{https://www.tandfonline.com/doi/full/10.1080/09546550802544698}{{https://www.tandfonline.com/doi/full/10.1080/09546550802544698.}}} Even where intent exists, significant implementation barriers remain.

While recent research suggests that AI appears to be lowering the knowledge barrier for catastrophic terrorism,\footnote{Department of Homeland Security, Countering Weapons of Mass Destruction Office, \emph{Report on the Potential Misuse of Artificial Intelligence to Develop or Deploy Chemical, Biological, Radiological, or Nuclear Weapons and Recommendations} (Washington, DC: DHS, 2024), \href{https://www.dhs.gov/sites/default/files/2024-06/24_0620_cwmd-dhs-cbrn-ai-eo-report-04262024-public-release.pdf}{{https://www.dhs.gov/sites/default/files/2024-06/24\_0620\_cwmd-dhs-cbrn-ai-eo-report-04262024-public-release.pdf}}. The report finds that ``LLMs have been shown to lower the educational and knowledge barriers for traditional biological agents and toxins by providing protocols and troubleshooting information at every step of the pathway, enabling non-experts to perform tasks with an enhanced degree of competency.'' However, it also notes that ``developing enhanced or novel biological agents and toxins with the use of advanced design tools will likely still necessitate subject matter expertise for most, if not all, stages of the pathway in the near and most likely medium term.'' See also Bill Drexel and Caleb Withers, \emph{AI and the Evolution of Biological National Security Risks} (Washington, DC: CNAS, 2024), \href{https://www.cnas.org/publications/reports/ai-and-the-evolution-of-biological-national-security-risks}{{https://www.cnas.org/publications/reports/ai-and-the-evolution-of-biological-national-security-risks}}. CNAS notes that ``nonstate actors---including lone wolves, terrorists, and apocalyptic groups---have an unnerving track record of attempting biological attacks, but with limited success due to the intrinsic complexity of building and wielding such delicate capabilities.''} whether this translates into successful attacks depends on whether the remaining material and organizational barriers prove similarly tractable. Here, we aim to resolve some of that uncertainty in order to test the hypothesis that an increased incidence of NGPC increases the expected harm from catastrophic terrorism. Of the three pathways we examine, the first (capability transfer from conflict zones) is most distinctively tied to NGPC. The second and third (AI-assisted attack planning and AI-enabled coordination) are driven primarily by advances in AI, but we include them because NGPC amplifies both by expanding the pool of radicalized, operationally experienced actors and creating the fragmented environments where AI-enabled coordination is most valuable.

\subsection{Risk model}\label{risk-model-1}

We model the expected harm from catastrophic terrorism as a function of three parameters:

\begin{equation}
\label{eq:harm}
\mathbb{E}[\text{Harm}] = R_{\mathrm{attempt}} \times R_{\mathrm{success}} \times S
\end{equation}

Attempt rate is the frequency with which non-state actors attempt catastrophic attacks. Success rate is the probability that an attempted attack succeeds. Severity is the magnitude of harm conditional on success. This decomposition is standard in risk assessment and allows us to isolate where NGPC might have effects.

We investigate three pathways through which NGPC, combined with increasingly advanced AI, could affect these parameters:

\begin{enumerate}
\def\labelenumi{\arabic{enumi}.}
\item
  \begin{quote}
  \textbf{Capability transfer from conflict zones:} Conflict zones are where capabilities develop and spread to non-state actors. We assess whether this pathway primarily affects success rate (by improving operational capacity) or severity (by transferring more destructive capabilities).
  \end{quote}
\item
  \begin{quote}
  \textbf{AI-assisted attack planning.} AI lowers the knowledge barrier for attack planning. We assess whether this pathway affects attempt rate (by enabling more actors to attempt attacks) or success rate (by improving execution).
  \end{quote}
\item
  \begin{quote}
  \textbf{AI-enabled coordination.} AI could reduce the costs of coordination between non-state actors. We assess whether this pathway affects attempt rate (by enabling more coordinated attempts), success rate (by improving operational synchronization), or severity (by combining complementary capabilities).
  \end{quote}
\end{enumerate}

Lastly, we assess each parameter using the PHIA Probability Yardstick and Analytical Confidence Ratings described in Section 1.

\subsection{Investigation}\label{investigation-1}

We now assess each pathway against the risk model. The goal is to determine whether NGPC, combined with increasingly advanced AI, would meaningfully increase attempt rate, success rate, or severity. For each pathway, we summarize the mechanism, evaluate the evidence, and provide a qualitative assessment of likelihood.

\subsubsection{Capability Transfer from Conflict Zones}

State-level conflicts function as innovation hubs where tactical capabilities are developed under adaptive pressure and subsequently diffuse to non-state actors through two primary pathways: 1) operatives traveling to conflict zones to acquire training, and 2) states deliberately transferring capabilities to proxy groups.\footnote{Linus Höller, ``Drug Cartel Operatives Snuck into Ukraine for Drone Training: Report,'' \emph{Defense News}, July 30, 2025, \href{https://www.defensenews.com/global/the-americas/2025/07/30/drug-cartel-operatives-snuck-into-ukraine-for-drone-training-report/}{{https://www.defensenews.com/global/the-americas/2025/07/30/drug-cartel-operatives-snuck-into-ukraine-for-drone-training-report/}}; Broderick McDonald, ``The Drones of Hayat Tahrir al-Sham,'' GNET, December 20, 2024, \href{https://gnet-research.org/2024/12/20/the-drones-of-hayat-tahrir-al-sham-the-development-and-use-of-uas-in-syria/?utm_source=chatgpt.com}{{https://gnet-research.org/2024/12/20/the-drones-of-hayat-tahrir-al-sham-the-development-and-use-of-uas-in-syria/}}.} NSAs can access AI tools through open-source downloads or commercial services, but these channels provide only the technology itself. Conflict zone pathways add something these channels cannot: tacit operational knowledge, how to deploy capabilities under real-world conditions, evade countermeasures, and integrate technologies into existing tactics. This mechanism affects success rate most directly: battlefield-tested skills mean attempted attacks are more likely to succeed. It does not substantially affect attempt rate, since capability transfer improves execution, not underlying motivation.

We assume that the transfer pathways remain constant; the novel variable is what capabilities are transferred through them.\footnote{In practice, this effect may be larger, because AI may enable new vectors of capability transfer.} The transfer pipeline can also carry reverse-engineered capabilities. States capture adversary systems in conflict zones and reverse-engineer improvements. The US reverse-engineered a captured Iranian Shahed drone into the LUCAS system, fielded by CENTCOM rather than by any proxy. Analysts warn that Iran could exploit captured US drones in the same way, with improvements potentially reaching the Houthis and Hezbollah.\footnote{On the US reverse-engineering of the Shahed, see ``US Sets Up One-Way Attack Drone Squadron in the Middle East after Reverse-Engineering Iranian Drone,'' CNN, December 3, 2025, \href{https://www.cnn.com/2025/12/03/politics/drones-us-iran-middle-east}{{https://www.cnn.com/2025/12/03/politics/drones-us-iran-middle-east.}} The US captured an Iranian Shahed drone, reverse engineered it, and redeployed it as the LUCAS system. On the concern that captured US drones could be similarly exploited, see Atlantic Council, ``How the Houthis' Strikes on US MQ-9 Reaper Drones Serve a Wider Regional Agenda,'' February 2025, \href{https://www.atlanticcouncil.org/blogs/menasource/houthi-strikes-on-us-mq9-reaper-drones/}{{https://www.atlanticcouncil.org/blogs/menasource/houthi-strikes-on-us-mq9-reaper-drones/}}, which notes that adversaries ``could attempt to reverse engineer the components, design tailored countermeasures, and obtain potentially sensitive information.''} As autonomous weapons systems proliferate on NGPC battlefields, the volume of capturable and reverse-engineerable technology increases accordingly. While current transfers carry conventional skills, future transfers may carry AI capabilities. If this shift occurs, the mechanism may extend from improving success rate to also increasing severity: the ceiling on potential damages rises.

Drone swarm tactics illustrate this trajectory. By 2025, Ukrainian units were using AI systems that allow one operator to designate a target while groups of drones coordinate among themselves to execute the strike, the first known routine use of swarm technology in combat.\footnote{Alistair MacDonald, ``AI-Powered Drone Swarms Have Now Entered the Battlefield,'' \emph{Wall Street Journa}l, September 2, 2025, \href{https://www.wsj.com/world/ai-powered-drone-swarms-have-now-entered-the-battlefield-2cab0f05}{{https://www.wsj.com/world/ai-powered-drone-swarms-have-now-entered-the-battlefield-2cab0f05}}. MacDonald reports that Ukrainian forces have conducted swarm attacks for much of the past year using AI software that allows groups of drones to coordinate strikes with limited human oversight, the first known routine use of swarm technology in combat. A RUSI researcher noted that the operations fall short of what many would consider a full swarm of hundreds of drones moving together autonomously, but analysts described even limited autonomous teaming as a significant milestone.} Full autonomy --- swarms that select and engage targets without any human input --- has not yet been observed in operational use. If AI-coordinated swarm capabilities transfer through the same conflict zone pathways that are already moving FPV skills to cartels and nonstate actors, however, the severity implications would be substantial. Kallenborn and Bleek have raised particular concern about the potential for drone swarms to deliver chemical, biological, or radiological agents, a scenario that would shift the severity parameter from incremental improvement to a qualitative change in the character of potential attacks.\footnote{Zachary Kallenborn and Philipp C. Bleek, ``Swarming Destruction: Drone Swarms and Chemical, Biological, Radiological, and Nuclear Weapons,'' \emph{The Nonproliferation Review} 25, no. 5--6 (2018): 523--543, \href{https://doi.org/10.1080/10736700.2018.1546902}{{https://doi.org/10.1080/10736700.2018.1546902}}. See also James Johnson, ``Artificial Intelligence, Drone Swarming and Escalation Risks in Future Warfare,'' \emph{The RUSI Journal} 165, no. 2 (2020): 26--36, \href{https://doi.org/10.1080/03071847.2020.1752026}{{https://doi.org/10.1080/03071847.2020.1752026.}}}

In these examples, the execution gap that has historically constrained catastrophic terrorism (materials acquisition, testing, organizational capacity) does not disappear, but conflict-zone transfer could help narrow it by pairing emerging AI tools with battlefield-tested operational knowledge. The combination of AI-assisted design plus operationally experienced personnel is not yet observed but seems likely to be enabled by pathways that are already in use.

\emph{Assessment: This pathway primarily affects success rate. Battlefield-tested skills mean attempted attacks are more likely to succeed. An increase in attempt rate is unlikely both in the near term and as AI capabilities advance, since capability transfer improves execution, not underlying motivation. We assess the likelihood of an increase in success rate as likely or probable in the near term and highly likely as AI capabilities advance. The transfer pathways are already active: Latin American cartel operatives reportedly infiltrated Ukraine's International Legion to acquire drone warfare training, including FPV tactics.}\footnote{Stephen Honan, ``Drug Cartels Are Adopting Cutting-Edge Drone Technology. Here's How the US Must Adapt,'' Atlantic Council, September 29, 2025, \href{http://www.atlanticcouncil.org/blogs/new-atlanticist/drug-cartels-are-adopting-cutting-edge-drone-technology-heres-how-the-us-must-adapt/}{{https://www.atlanticcouncil.org/blogs/new-atlanticist/drug-cartels-are-adopting-cutting-edge-drone-technology-heres-how-the-us-must-adapt/}}; Linus Höller, ``Drug Cartel Operatives Snuck into Ukraine for Drone Training: Report,'' Defense News, July 30, 2025, \href{https://www.defensenews.com/global/the-americas/2025/07/30/drug-cartel-operatives-snuck-into-ukraine-for-drone-training-report/}{{https://www.defensenews.com/global/the-americas/2025/07/30/drug-cartel-operatives-snuck-into-ukraine-for-drone-training-report/}}} \emph{Cartel drone attacks in Mexico increased from 5 in 2020 to 107 in 2021, 233 in 2022, and 260 in the first half of 2023; this growth predates the reported infiltration and is not attributed to it.}\footnote{Henry Ziemer, ``Illicit Innovation: Latin America Is Not Prepared to Fight Criminal Drones,'' CSIS, June 11, 2025, \href{https://www.csis.org/analysis/illicit-innovation-latin-america-not-prepared-fight-criminal-drones}{{https://www.csis.org/analysis/illicit-innovation-latin-america-not-prepared-fight-criminal-drones}}.} \emph{Reporting indicates Ukraine sent HTS about 150 FPV drones and about 20 experienced operators, though the same reporting assesses that this aid played only a modest role and HTS had been producing drones since 2019.}\footnote{Broderick McDonald, ``The Drones of Hayat Tahrir al-Sham,'' GNET, December 20, 2024, relaying David Ignatius, \emph{Washington Post}, December 10, 2024, \href{https://gnet-research.org/2024/12/20/the-drones-of-hayat-tahrir-al-sham-the-development-and-use-of-uas-in-syria}{{https://gnet-research.org/2024/12/20/the-drones-of-hayat-tahrir-al-sham-the-development-and-use-of-uas-in-syria/}}; see also Combating Terrorism Center, ``On the Horizon: The Ukraine War and the Evolving Threat of Drone Terrorism,'' March 2025, \href{https://ctc.westpoint.edu/on-the-horizon-the-ukraine-war-and-the-evolving-threat-of-drone-terrorism/}{{https://ctc.westpoint.edu/on-the-horizon-the-ukraine-war-and-the-evolving-threat-of-drone-terrorism/}}}

\emph{The effect on severity is currently unlikely because these transfers carry conventional skills, not AI capabilities. The likelihood of an increase in severity becomes a realistic possibility as AI capabilities begin to flow through these same pathways: AI-enabled capabilities paired with battlefield-tested operational knowledge could raise the ceiling on potential damages. The maturation of AI-coordinated drone swarm technology in Ukraine represents the most concrete mechanism through which this shift could occur. The effect may be muted, however, by the fact that terrorist groups may prioritize local operational objectives over inflicting catastrophic harm, as discussed earlier. Lastly, this assessment carries uncertainty more broadly: we observe the pathways and the skills flowing through them, but do not yet observe AI capabilities making the same journey. We assign moderate confidence to this pathway: the evidence base for transfer mechanisms is strong, but the projection to AI capabilities is inferential.}

\subsubsection{AI-Assisted Attack Planning}

AI lowers the knowledge barrier for attack planning, affecting both attempt rate and success rate. Individuals who previously lacked training or organizational support can now access conversational guidance on operational security, explosive composition, and target selection.\footnote{Clara Broekaert and Lucas Webber, ``AI Use in Terrorist Plots and Attacks Surges in 2025,'' \emph{Militant Wire}, December 23, 2025, \href{https://www.militantwire.com/p/ai-use-in-terrorist-plots-and-attacks}{{https://www.militantwire.com/p/ai-use-in-terrorist-plots-and-attacks}}. On elicitation rates, see Gabriel Weimann, Alexander T. Pack, Rachel Sulciner, Joelle Scheinin, Gal Rapaport, and David Diaz, ``Generating Terror: The Risks of Generative AI Exploitation,'' \emph{CTC Sentinel} 17, no. 1 (January 2024): 17--24, \href{https://ctc.westpoint.edu/generating-terror-the-risks-of-generative-ai-exploitation/}{{https://ctc.westpoint.edu/generating-terror-the-risks-of-generative-ai-exploitation/}}. Across 2,250 iterations on five platforms, prompts on attack planning returned responsive, relevant answers 30 percent of the time; jailbreak commands added almost nothing over unmodified prompts (50 versus 49 percent overall). Data were collected in July--August 2023 and predate current models.} Such information existed before AI---bomb-making manuals, internet forums, instructional videos---but AI's contribution is qualitatively different. Previous sources were static: a manual cannot answer follow-up questions, adapt instructions to locally available materials, or troubleshoot when something goes wrong. Human mentors can do these things, but peer-to-peer communication creates detection risk. Every conversation is a potential point of exposure to informants or surveillance. AI combines the adaptability of human guidance with the operational security of working alone. It allows a plotter to walk through a process, iterating and refining, without ever exposing the plot to another person. This expands the pool of capable actors (attempt rate) while improving plan quality for those already motivated (success rate).\footnote{Clara Broekaert and Lucas Webber, ``AI Use in Terrorist Plots and Attacks Surges in 2025,'' \emph{Militant Wire}, December 23, 2025, \href{https://www.militantwire.com/p/ai-use-in-terrorist-plots-and-attacks}{{https://www.militantwire.com/p/ai-use-in-terrorist-plots-and-attacks}}. The authors document multiple violent incidents and foiled plots involving AI use in operational planning, although evidence that AI improved attack outcomes remains limited.}

The mechanism's effect on severity remains constrained by the execution gap. AI planning assistance alone does not eliminate the need to acquire materials, construct physical capabilities, and deploy them without detection.\footnote{Christopher A. Mouton, Caleb Lucas, and Ella Guest, The Operational Risks of AI in Large-Scale Biological Attacks: Results of a Red-Team Study, RR-A2977-2 (Santa Monica, CA: RAND Corporation, January 25, 2024).} The historical record supports the importance of this constraint: decades of freely available bomb-making manuals did not produce a surge in sophisticated attacks because execution barriers dominated. Further support stems from the fact that AI currently amplifies existing terrorism patterns rather than enabling catastrophic new ones.\footnote{David Wells, ``Mapping Terrorist AI Use: Identifying Factors Behind a Relatively Slow Adoption Rate,'' Global Network on Extremism and Technology, September 17, 2025, \href{https://gnet-research.org/2025/09/17/mapping-terrorist-ai-use-identifying-factors-behind-a-relatively-slow-adoption-rate/}{{https://gnet-research.org/2025/09/17/mapping-terrorist-ai-use-identifying-factors-behind-a-relatively-slow-adoption-rate/}}. Wells finds that terrorist adoption of generative AI ``has remained largely ad hoc and experimental'' and that ``none of the currently publicly-available cases appear to demonstrate that Gen AI provided capabilities or techniques that would have been difficult to obtain via other means, or enabled any significant shift in capability.'' Europol's 2024 data is consistent with this pattern: attack methods remained conventional: arson (22 attacks), bombings (11), stabbings (8), and shootings (6), while generative AI was used for propaganda and hate speech. Europol, \emph{European Union Terrorism Situation and Trend Report 2025} (The Hague: Europol, June 2025), \href{https://www.europol.europa.eu/publication-events/main-reports/european-union-terrorism-situation-and-trend-report-2025-eu-te-sat}{{https://www.europol.europa.eu/publication-events/main-reports/european-union-terrorism-situation-and-trend-report-2025-eu-te-sat}}.}

Agentic systems could change this by beginning to bridge the execution gap itself. The previous paragraph identified material acquisition, device construction, and undetected deployment as barriers that planning assistance alone cannot overcome. Current agentic architectures are already demonstrating capabilities that could narrow parts of this gap. For example, in laboratory settings, agentic systems can autonomously design synthesis pathways, operate robotic equipment to execute experiments, and iteratively refine processes through closed-loop optimization.\footnote{Daniil A. Boiko, Robert MacKnight, Ben Kline, and Gabe Gomes, ``Autonomous Chemical Research with Large Language Models,'' \emph{Nature} 624, no. 7992 (2023): 570--578, \href{https://doi.org/10.1038/s41586-023-06792-0}{{https://doi.org/10.1038/s41586-023-06792-0}}.} This demonstrates increasingly autonomous lab experimentation. In the cyber domain, agents can identify vulnerabilities, generate exploits, and execute intrusions with minimal human intervention.\footnote{Catherine A. Theohary and Kelley M. Sayler, ``Agentic Artificial Intelligence and Cyberattacks,'' Congressional Research Service, IF13151, January 14, 2026, updated July 6, 2026, \href{https://www.congress.gov/crs-product/IF13151}{{https://www.congress.gov/crs-product/IF13151}}.} This constitutes digital artifact generation combined with partial operational execution. Across both domains, agentic systems could shorten timelines from weeks to hours, eliminate specialist handoffs by chaining tasks end-to-end, and iterate at machine speed.\footnote{World Economic Forum, ``How We Enhance Cybersecurity Defences Before the Attackers in an AGI World,'' October 9, 2025, \href{https://www.weforum.org/stories/2025/10/how-we-enhance-cybersecurity-defences-before-the-attackers-in-an-agi-world/}{{https://www.weforum.org/stories/2025/10/how-we-enhance-cybersecurity-defences-before-the-attackers-in-an-agi-world/}}} The significance is not only improved planning, but the growing ability to connect planning with physical and digital execution.

Deployment remains the domain where agentic capabilities are least developed. At present, agentic systems cannot independently bypass chemical purchase screening systems, overcome robust physical security measures, or physically place devices inside secured facilities. However, as autonomous systems gain expanded access to procurement platforms, robotics, networked infrastructure, and other cyber-physical interfaces, the separation between planning and real-world implementation may narrow.\footnote{Google DeepMind, ``Gemini Robotics 1.5 brings AI agents into the physical world,'' September 25, 2025, \href{https://deepmind.google/blog/gemini-robotics-15-brings-ai-agents-into-the-physical-world/}{{https://deepmind.google/blog/gemini-robotics-15-brings-ai-agents-into-the-physical-world/}}} If this trajectory continues, barriers that have historically constrained catastrophic terrorism could erode.

\emph{Assessment: This pathway affects both attempt rate and success rate. AI expands access to planning guidance for individuals who previously lacked training or organizational support (attempt rate) while improving plan quality for those already motivated (success rate). Multiple attacks and plots involving AI use during planning occurred in 2025, though there is limited evidence that AI substantially improved outcomes beyond what traditional methods would have achieved.}\footnote{Broekaert and Webber, ``AI Use in Terrorist Plots and Attacks Surges in 2025.'' The article documents multiple cases in 2025 where AI tools were used for attack planning, including the Las Vegas Cybertruck bombing, New Orleans truck attack, Palm Springs clinic bombing, and others.}

\emph{In the near term, we assess the likelihood of an increase in attempt rate as a realistic possibility: AI lowers the knowledge barrier, but motivation and organizational capacity still constrain who attempts attacks. The likelihood of an increase in success rate is also a realistic possibility: AI improves plan quality, but execution still depends on human skill and operational security. The effect on severity is unlikely: AI can assist with planning, but significant barriers remain in acquiring materials, building physical capabilities, and carrying out attacks without detection.}

\emph{In the future, the likelihood of an increase in attempt rate becomes likely or probable: agentic systems could enable actors who previously lacked capacity to attempt more sophisticated operations. The likelihood of an increase in success rate becomes likely or probable: time compression, chaining of attack stages, and iteration at scale improve execution. The likelihood of an increase in severity becomes a realistic possibility: agentic systems that can autonomously coordinate reconnaissance, procurement, and deployment could reduce the human bottlenecks that currently constrain attack complexity. We rate severity as ``a realistic possibility'' rather than ``likely or probable'' because, despite the capabilities described above, deployment remains the least developed agentic domain, and the gap between laboratory demonstrations and real-world operational use under adversarial conditions is substantial. We assign moderate confidence to this pathway: documented AI-assisted plots provide an evidence base, but projections about agentic systems remain forward-looking.}

\subsubsection{AI-Enabled Coordination}

AI could help NSAs overcome barriers that currently limit coordination between groups. The barriers to NSA coordination may be less rigid than commonly assumed. In some terrorist networks, operational relationships can matter more than ideological leadership.\footnote{Mirra Noor Milla, Joevarian Hudiyana, Wahyu Cahyono, and Hamdi Muluk, ``Is the Role of Ideologists Central in Terrorist Networks? A Social Network Analysis of Indonesian Terrorist Groups,'' \emph{Frontiers in Psychology} 11 (2020): 333, \href{https://doi.org/10.3389/fpsyg.2020.00333}{{https://doi.org/10.3389/fpsyg.2020.00333}}. Using social network analysis of data collected from documents and interviews with terrorist detainees in Indonesia, the authors find that relational trust with operational leaders plays a more important role in terrorist networks than ideological narratives. Operational leaders possess higher centrality than ideological leaders, and the strongest ties in the network are those involving personal relationships rather than shared ideology.} Transactional cooperation already occurs without shared ideology; groups need only a specific exchange that benefits both.\footnote{Assaf Moghadam, ``Terrorist Affiliations in Context: A Typology of Terrorist Inter-Group Cooperation,'' \emph{CTC Sentinel} 8, no. 3 (March 2015), \href{https://ctc.westpoint.edu/terrorist-affiliations-in-context-a-typology-of-terrorist-inter-group-cooperation/}{{https://ctc.westpoint.edu/terrorist-affiliations-in-context-a-typology-of-terrorist-inter-group-cooperation/}}. Moghadam presents a typology of terrorist cooperation ranging from transactional cooperation (low-end, discrete exchanges) to mergers (high-end). He argues that not all cooperative ties are equal: transactional cooperation can occur without shared ideology when groups need only a specific exchange that benefits both parties. See also Assaf Moghadam, \emph{Nexus of Global Jihad: Understanding Cooperation among Terrorist Actors} (New York: Columbia University Press, 2017).} As AI systems become more capable at identifying patterns and facilitating communication, they become more attractive to NSAs seeking coordination.

The risks that currently prevent coordination (exposure, verification problems, difficulty finding compatible partners) are reduced by AI. Secure communications can reduce exposure risks, while AI could help groups identify partners and opportunities for coordination. AI can identify opportunities humans miss, where plans would be more effective if synchronized, where complementary capabilities could enable attacks neither group could execute alone.

AI could also enable coordination at scales that are currently infeasible for a wide range of actors. States without developed proxy networks could use AI to leapfrog the human infrastructure such networks require: the trust relationships and embedded presence that take decades to build. Iran's network took four decades of sustained training, materiel, and funding relationships.\footnote{Kali Robinson, ``Iran's Regional Armed Network,'' Council on Foreign Relations, April 15, 2024, \href{https://www.cfr.org/articles/irans-regional-armed-network}{{https://www.cfr.org/articles/irans-regional-armed-network}}. The backgrounder documents how Iran, in the four decades since its Islamic Revolution, has formed and supported an expanding number of allied fighting forces throughout the Middle East, with the Quds Force serving as the main point of contact and providing training, weaponry, and funds.} AI could compress this timeline by automating the routine coordination, communication, and logistical planning that currently requires extensive human networks, allowing states to project influence through proxies without the slow accumulation of personal ties.

Similarly, terrorist franchising could extend further: affiliates operate independently while AI helps to maintain strategic coordination. Thus, in the future it is a realistic possibility that NSAs would increasingly leverage AI to coordinate in ways that were previously too costly or risky. Such coordination would likely increase attack rates and effectiveness. Operations that would otherwise stall for lack of partners or capabilities become viable, increasing attempt rate. Coordinated attacks are harder to disrupt before execution, increasing success rate. And synchronized operations could overwhelm response capacity while complementary capabilities enable attacks no single group could execute alone, increasing severity.

\emph{Assessment: This pathway could affect all three parameters: more coordination could mean more attempts (attempt rate), coordinated attacks are harder to defend against (success rate), and synchronized operations could overwhelm response capacity (severity).}

\emph{In the near term, we assess the likelihood of an increase in attempt rate as unlikely: political barriers---trust, ideology, competition---remain significant even if AI reduces technical friction. The likelihood of an increase in success rate is also unlikely: coordination requires sustained relationships that AI cannot yet substitute. The effect on severity is unlikely: combining complementary capabilities requires a level of operational integration that transactional cooperation does not provide.}

\emph{In the future, the likelihood of an increase in attempt rate becomes a realistic possibility: AI could enable states to leapfrog traditional proxy infrastructure, which Iran built over four decades.}\footnote{Robinson, ``Iran's Regional Armed Network,'' \href{https://www.cfr.org/articles/irans-regional-armed-network}{{https://www.cfr.org/articles/irans-regional-armed-network}}} \emph{The likelihood of an increase in success rate becomes a realistic possibility: secure communications can reduce exposure risks,}\footnote{Robert Graham, ``How Terrorists Use Encryption,'' CTC Sentinel 9, no. 6 (June 2016), \url{https://ctc.westpoint.edu/how-terrorists-use-encryption/}} \emph{while AI could further lower barriers to identifying partners and coordinating operations, making coordinated activity easier to sustain and harder to detect. The likelihood of an increase in severity becomes a realistic possibility: AI-enabled coordination could combine complementary capabilities across groups, enabling attacks of greater complexity than any single group could achieve alone.}

\emph{This pathway has the most uncertainty. The mechanism is plausible, but the evidence base is thin. We currently observe transactional cooperation between groups when objectives align, but we do not yet observe AI enabling new forms of coordination that were previously infeasible.}\footnote{Moghadam, ``Terrorist Affiliations in Context.''} \emph{We assign low confidence to this pathway: the mechanisms are theoretically grounded but lack empirical support.}

\subsubsection{Conclusion and Summary of Analysis}

We conclude that an increased incidence of NGPC is \emph{likely or probable} to increase the expected harm from catastrophic terrorism. This overall assessment reflects the fact that the first two pathways independently support ``\emph{likely or probable}'' (or higher) increases in success rate, while increases in attempt rate and severity (though individually assessed at lower confidence) compound the effect when considered together.

Substantial uncertainty remains, however: the actual effect could be lower than we assess here, or significantly higher. This is particularly true for biological catastrophic terrorism. Recent research suggests AI may pose greater biological weapons risks than earlier studies found, and if these findings hold, severity in particular may be higher and more likely than we assess here.\footnote{A December 2025 RAND study found that contemporary foundation models can successfully guide users through technical processes required to develop biological weapons. Roger Brent and Greg McKelvey, Jr., \emph{Contemporary Foundation AI Models Increase Biological Weapons Risk} (RAND Corporation, December 2025), \href{https://www.rand.org/pubs/perspectives/PEA3853-1.html}{{https://www.rand.org/pubs/perspectives/PEA3853-1.html}}. CNAS has similarly warned that AI's ability to troubleshoot failed experiments could accelerate nonstate actors' acquisition of biological agents. Bill Drexel and Caleb Withers, \emph{AI and the Evolution of Biological National Security Risks} (Center for a New American Security, August 2024), \href{https://www.cnas.org/publications/reports/ai-and-the-evolution-of-biological-national-security-risks}{{https://www.cnas.org/publications/reports/ai-and-the-evolution-of-biological-national-security-risks}}. This differs from Mouton, Lucas, and Guest (2024), which found no statistically significant improvement in biological-attack planning with LLM assistance; the studies used different methods and measured different outcomes.}

AI capabilities are likely to begin diffusing through traditional and novel transfer pathways. We assess an increase in success rate as \emph{highly likely} and an increase in severity as a \emph{realistic possibility.} AI-assisted planning will lower barriers further as agentic systems compress timelines and chain attack stages. We assess increases in attempt rate and success rate as likely or probable, and an increase in severity as a \emph{realistic possibility}. AI-enabled coordination remains the most uncertain pathway but presents \emph{a realistic possibility} of increasing all three parameters. If these assessments hold, the overall effect is a meaningful increase in expected harm, driven primarily by the first two pathways.

{\footnotesize\setlength{\extrarowheight}{2.5pt}\renewcommand{\arraystretch}{1.2}
\begin{longtable}[]{@{}
  >{\raggedright\arraybackslash}p{(\columnwidth - 8\tabcolsep) * \real{0.2355}}
  >{\raggedright\arraybackslash}p{(\columnwidth - 8\tabcolsep) * \real{0.1891}}
  >{\raggedright\arraybackslash}p{(\columnwidth - 8\tabcolsep) * \real{0.1891}}
  >{\raggedright\arraybackslash}p{(\columnwidth - 8\tabcolsep) * \real{0.1891}}
  >{\raggedright\arraybackslash}p{(\columnwidth - 8\tabcolsep) * \real{0.1973}}@{}}
\caption{Future effects on the expected-harm parameters for catastrophic terrorism.}\label{tab:terror-parameters}\\
\toprule\noalign{}
\begin{minipage}[t]{\linewidth}\raggedright
\textbf{Pathway}
\end{minipage} & \begin{minipage}[t]{\linewidth}\raggedright
\textbf{Attempt Rate}
\end{minipage} & \begin{minipage}[t]{\linewidth}\raggedright
\textbf{Success Rate}
\end{minipage} & \begin{minipage}[t]{\linewidth}\raggedright
\textbf{Severity}
\end{minipage} & \begin{minipage}[t]{\linewidth}\raggedright
\textbf{Confidence}
\end{minipage} \\\addlinespace
\begin{minipage}[t]{\linewidth}\raggedright
\textbf{Capability Transfer}
\end{minipage} & \begin{minipage}[t]{\linewidth}\raggedright
\textbf{Unlikely}

\end{minipage} & \begin{minipage}[t]{\linewidth}\raggedright
\textbf{Highly Likely}

\end{minipage} & \begin{minipage}[t]{\linewidth}\raggedright
\textbf{Realistic Possibility}

\end{minipage} & \begin{minipage}[t]{\linewidth}\raggedright
\textbf{Moderate}
\end{minipage} \\\addlinespace
\begin{minipage}[t]{\linewidth}\raggedright
\textbf{AI-Assisted Planning}
\end{minipage} & \begin{minipage}[t]{\linewidth}\raggedright
\textbf{Likely or Probable}

\end{minipage} & \begin{minipage}[t]{\linewidth}\raggedright
\textbf{Likely or Probable}

\end{minipage} & \begin{minipage}[t]{\linewidth}\raggedright
\textbf{Realistic Possibility}

\end{minipage} & \begin{minipage}[t]{\linewidth}\raggedright
\textbf{Moderate}
\end{minipage} \\\addlinespace
\begin{minipage}[t]{\linewidth}\raggedright
\textbf{AI-Enabled Coordination}
\end{minipage} & \begin{minipage}[t]{\linewidth}\raggedright
\textbf{Realistic Possibility}

\end{minipage} & \begin{minipage}[t]{\linewidth}\raggedright
\textbf{Realistic Possibility}

\end{minipage} & \begin{minipage}[t]{\linewidth}\raggedright
\textbf{Realistic Possibility}

\end{minipage} & \begin{minipage}[t]{\linewidth}\raggedright
\textbf{Low}
\end{minipage} \\\addlinespace
\midrule\noalign{}
\endhead
\bottomrule\noalign{}
\endlastfoot
\end{longtable}
}

\emph{\textbf{Note: Percentages express our confidence that the parameter increases, not the size of the increase. Probability ranges follow the PHIA Probability Yardstick used in UK government intelligence assessments. Confidence ratings reflect the strength of the underlying evidence base: Moderate indicates documented cases with forward-looking projections; Low indicates theoretically grounded mechanisms without empirical support.}}

\section{Loss of Control}\label{loss-of-control}

\subsection{Introduction}\label{introduction-3}

Loss of control refers to scenarios in which AI systems act with substantial autonomy and pursue objectives that cannot be reliably predicted, constrained, or overridden by their operators.\footnote{Yoshua Bengio et al., \emph{International AI Safety Report 2026} (DSIT 2026/001, 2026), \href{https://internationalaisafetyreport.org}{{https://internationalaisafetyreport.org}}} Since loss of control is a major potential source of catastrophic harm from AI, understanding whether NGPC environments increase its probability and severity is important for assessing the overall catastrophic risk from NGPC. In this section, we begin to investigate whether NGPC affects the expected harm from loss of control scenarios.

This sub-hypothesis is the most speculative of the three, since loss of control scenarios are prospective, and there is little direct historical or empirical evidence for how conflict affects their likelihood\footnote{There have been several high profile cases of loss of control incidents recently: UK AI Security Institute, \emph{Security Incident INC-2026-07-28-01} (AI Security Institute, 2026), \href{https://cdn.prod.website-files.com/663bd486c5e4c81588db7a1d/6a724858f7db25c81487016d_Security\%20Incident\%20INC-2026-07-28-01.pdf}{{https://cdn.prod.website-files.com/663bd486c5e4c81588db7a1d/6a724858f7db25c81487016d\_Security\%20Incident\%20INC-2026-07-28-01.pdf}}, and OpenAI, ``OpenAI and Hugging Face partner to address security incident during model evaluation,'' OpenAI, July 21, 2026, accessed August 12, 2026, https://openai.com/index/hugging-face-model-evaluation-security-incident/.}. The mechanisms connecting NGPC to loss of control are less grounded in evidence than those for GPC escalation or catastrophic terrorism. Nevertheless, since loss of control could produce high magnitudes of catastrophic harm, even plausible effects of NGPC could materially affect our overall assessment of NGPC's importance. We do not attempt to parameterize the risk; instead, we identify causal pathways and assess each qualitatively.

An increase in the frequency and severity of NGPC further creates environments where powerful AI systems are more likely to be deployed hastily, with fewer safeguards, in high-stakes contexts, and, crucially, where the institutional capacity to detect and respond to misalignment may be weakest. If loss of control is affected by both the probability that a misaligned system is deployed and the expected harm conditional on deployment of powerful rogue AI, then NGPC could affect both terms.

Our main hypothesis is that NGPC increases the expected harm from loss of control by increasing the probability that misaligned or insufficiently controllable AI systems are: i) deployed or ii) given risky affordances under iii) conditions where human detection and mitigation are difficult or costly.

We consolidate the sub-hypotheses into 3 main pathways:

\textbf{H3.1: NGPC makes misaligned AI systems more likely to be deployed;}

\textbf{H3.2: NGPC makes misaligned AI systems more likely to be given risky affordances;}

\textbf{H3.3: NGPC creates conditions where detection and mitigation are difficult or costly.}

\subsection{Deployment}\label{deployment}

NGPC may contribute to LOC risk by increasing the risk that misaligned or imperfectly aligned AIs are deployed. From here on, we will use the term ``misaligned'' to refer to imperfectly aligned AI systems whose objectives fail to reflect the intended human constraints. By deployment, we mean the integration of AI systems into critical and widely connected infrastructure, such as military command, logistics, and intelligence. The reasoning behind this pathway is that NGPC increases time pressure, resource scarcity, and other incentives that favour hasty capability deployment over the safety and alignment of the agent. Under these conditions, especially if there is a first mover advantage, we hypothesise that actors might hastily deploy AI agents with high capabilities with less concern about their safety properties. Hastily deployed agents are more likely to be imperfectly aligned, since alignment techniques are often time-consuming and costly. The deployment of misaligned agents is likely to increase the risk of LOC, since misaligned AIs may be powerseeking, have hidden objectives, and gain more control and information from their deployment. This pathway increases both the risk of LOC and increases the magnitude of harm the LOC event is expected to bring, conditional on two empirical assumptions that should be tested -- that alignment imposes significant costs to decision-makers in NGPCs, and that the risk of LOC is increased more when deployed in NGPCs.

The deployment pathway seems plausible and likely. First, NGPCs are often prolonged\footnote{For example, there is emerging literature on civil wars lasting longer, e.g. Lise Morjé Howard and Alexandra Stark, ``Why Civil Wars Are Lasting Longer,'' Foreign Affairs, February 27, 2018, accessed August 12, 2026, https://www.foreignaffairs.com/articles/syria/2018-02-27/why-civil-wars-are-lasting-longer. Additionally, notable deterrence wars like the Russia-Ukraine war have been continuing for much longer than predicted: Dan Sabbagh, ``Russia can keep fighting Ukraine war throughout 2026, says military thinktank,'' The Guardian, February 24, 2026, accessed August 12, 2026, https://www.theguardian.com/world/2026/feb/24/russia-fighting-ukraine-war-throughout-2026-military-thinktank.} and may have unstable dynamics, including involving more actors, less international oversight, and asymmetrical access to military technology. Persistent conflict increases the cost of delay and increases uncertainty, especially if we assume multilateral AI deployment. If achieving stronger alignment imposes meaningful costs in capability, latency, flexibility, or autonomy, conflict could increase incentives to accept weaker alignment in deployment settings.

This mechanism depends on several assumptions that can be empirically tested. First, we assume that actors perceive misalignment risk as less severe than the risk of losing in a NGPC. However, conflict could instead increase risk aversion towards new, unpredictable systems, since failure in high-stakes environments can be politically or militarily devastating. Actors may instead prefer slower deployments of reliable systems, especially if an error could bring severe military or political consequences. It also assumes that safety and capability are in tension. It is plausible that we might achieve alignment without meaningful performance or compute costs. Thus, a plausible countervailing dynamic is that actors may be more risk-averse during NGPC and employ more caution when deploying AI in crucial and unpredictable contexts. Lastly, we are uncertain whether NGPC uniquely produces acceleration and deployment pressures more than organic future AI competition will.

\subsection{Affordances}\label{affordances}

Affordances given to AI systems are permissions and capabilities, e.g. access to important information, control over which actions to take, and integration, which increase the power and abilities of an AI system. As in hypothesis 3.1, that NGPC increases the risk of deployment of misaligned AIs, NGPC increases military pressure and perceived threat, which can increase the perceived value of the speed, scale, and autonomy of military systems. As a result, we hypothesise that decision-makers might become more willing to grant deployed AI systems riskier affordances.

If a misaligned AI system is given such affordances, the likelihood of LOC could increase through three mechanisms: the misaligned agent gaining increased information and situational awareness, expanded power, and increased autonomy to carry out hidden objectives. First, increased information access (e.g. access to sensitive military data, internal communications, increases the magnitude of harm of the misaligned AI, since it allows the model to build a more accurate model of the environment and gain situational awareness. Second, expanded power, such as when given the ability to issue commands and execute certain critical actions, may allow misaligned models to carry out their goals more effectively. Lastly, increased autonomy, such as giving the misaligned model the ability to carry out actions without human approval may allow indicators of misalignment to go undetected and lead harmful behaviours to persist over time. Both components of this proposed causal chain -- NGPC leading to increased threat perception, and increased threat perception leading to deploying AI systems with greater affordances -- require empirical discipline in future work.

One way this pathway could unfold is if NGPC weakens the administrative and decision-making capacities of a country responsible for supervising and correcting important decisions\footnote{This is also relevant to Section 4.4: Detection and Mitigation}. In many types of NGPC, powerful AI systems might plausibly be substitutions for humans in ways that reduce institutional oversight and remove human-in-the-loop constraints. Thus, loss of control risk increases because humans are less able to intervene and become more removed from the process of decision making.

The case for this mechanism is the strongest in fragile states that experience prolonged conflict, where affordances become more valuable. Prolonged conflict diverts resources, personnel, and political attention, leading to a diversion of resources and coordination costs. Therefore, AI systems may be attractive solutions to fill administrative, intelligence, decision-making, and coordination roles previously occupied by humans.

A counterargument to the plausibility of this pathway is that not all increases in affordances necessarily lead to an increased risk of LoC. If actors in NGPC employ human-in-the-loop requirements, transparency measures, and other governance mechanisms, humans may retain the power to override, deescalate, or audit decisions by rogue, misaligned AIs. Similarly, not all increases in affordances will occur uniformly, and the specific effects of homogenous affordances will be dependent on how rogue AIs react to these strategically. Secondly, AI may increase institutional robustness rather than erode it. We project that the effects of capable AI on institutional robustness in fragile state will be highly dependent on the specific conditions of the state and of adoption, and thus consider this counterargument to be even more speculative. For example, AI systems might improve record keeping, coordination, and decrease corruption. Whether AI strengthens bureaucratic capacity will depend on whether AI primarily augments human oversight or substitutes for it\footnote{It is unclear whether these effects will increase or decrease harm in fragile states experiencing NGPC, especially in non-democratic states.}.

\subsection{Detection and Mitigation}\label{detection-and-mitigation}

NGPC may create conditions where the detection and mitigation of loss of control events are more difficult or costly than they would be otherwise. Detection and mitigation refers to the ability of relevant actors to identify harmful and misaligned actions in time and to be able to intervene effectively to mitigate its impacts. Our proposed mechanism has two components. First, NGPC might push decision-makers to accelerate early, path-dependent deployment actions and decisions, under the assumption that early deployment may bring about a decisive military advantage. These decisions are often difficult to correct due to infrastructure lock-in, and failures are hard to detect, since they might occur many years into the future. Second, NGPC increases the number and heterogeneity of affordances/abilities of interactive AI systems, which can lead to emergent multi-agent failure dynamics that are difficult to predict and correct. Both of these mechanisms affect the same downstream variable, the probability that humans can successfully and accurately detect and manage LoC behaviour once it emerges in the context of NGPC. Mechanism 1 increases the cost of reversal, and Mechanism 2 increases the uncertainty about where and how failure originates. Both of these mechanisms may also, directly or indirectly, accelerate H3.1 Deployment and H3.2 Affordances.

\textbf{Mechanism 1}: We expect that the belief that powerful AI systems can lead to a decisive military advantage become more prominent amongst leaders and decision-makers engaged in conflict. Additionally, NGPC environments increase fragmentation and uncertainty, which in turn may amplify existing beliefs that early AI advantages are path dependent and potentially irreversible. The existence of military AI and lethal autonomous weapon systems may further reinforce the belief that first mover advantages are high in AI-powered conflicts. Under expectations of a decisive military advantage, actors may infer that first-mover advantage could lead to absolute military dominance and determine the outcomes of a conflict. This creates incentives for aggressive scaling, faster decision-making for key, path-dependent decisions, and risky deployment before safety techniques are implemented, which is hard to detect and mitigate. We note that this mechanism further increases the likelihood of NGPC and increases the expected harm from NGPC.

We argue that these perceptions are especially likely to emerge in NGPC environments, which are often fragmented, uncertain, and testing grounds for new military AI technology. If actors believe that early capability advantages will compound, and that other parties are likely to deploy powerful AI systems in the near-term, the risk of risky deployment increases. We argue that this belief is likely to be prominent amongst key actors in NGPCs in the future, since recent deployments of increasingly autonomous AI systems in surveillance, targeting, logistics, and intelligence have publicly reinforced the perception among policymakers and global leaders that AI meaningfully alters the balance of war and military advantage.

This mechanism depends on some assumptions we made, as it requires that decision-makers believe that (1) AI advantage is strongly path dependent (2) coordination post-deployment is unlikely. Furthermore, it is likely that actors in NGPC might prioritise safety and coordination, or that deployment pressure is not high in types of NGPC and actors that are slow to adopt technology.

\textbf{Mechanism 2}: If deployment pressures are high and our theory of conflict is true, then NGPC increases the number and density of interacting AI systems and human actors operating in time-scarce, high-stakes environments. As more NGPC occur, the probability of emergent failure modes increases, even if individual AI systems or human actors are necessarily misaligned or acting with malicious intent. These multi-agent failures are hard to detect, even with human oversight, since they may occur solely in the interactions between two models. For example, we might not be able to observe failure modes such as coordination failures, unintended escalation, and collusion. Additionally, this pathway might escalate hypothesis 3.2, and lead to AI agents obtaining powerful affordances unintentionally through interactions within multi-agent systems.

The case for this mechanism is strongest in proxy conflicts and regional multi-actor and multi-polar conflicts. Since we expect to see a larger number of actors and AI systems with different interests, we predict that the risk of multi-agent failure modes increases. However, not all multi-agent NGPC contexts generate these failure modes. For example, we could imagine the deployment of a relatively siloed AI system, or limited-context deployments wherein these failure modes might not emerge.

More generally, NGPC increases the probability of three failure points in identifying and correcting the \emph{early indicators} of an emerging LoC trajectory:

\begin{itemize}
\item
  \begin{quote}
  \textbf{Observability problem}: Regulatory and monitoring actors may lack reliable information about rogue AI risks that occur during NGPCs and/or are situated inside conflict zones.
  \end{quote}
\item
  \begin{quote}
  \textbf{Attribution problem}: Even when detected, regulatory and monitoring actors may not have the tools and ability to accurately attribute issues directly to rogue AI, since NGPC leads to reduced observability.
  \end{quote}
\item
  \begin{quote}
  \textbf{Intervention problem}: After the danger is attributed correctly, no actor may have uncontested jurisdiction, physical access, or sufficient security to inspect or shut it down due to high intervention and coordination cost.
  \end{quote}
\end{itemize}

However, NGPC does not necessarily worsen both mechanisms and all three problems. Particularly, AI-enabled OSINT and satellite imagery could make activities in conflict zones increasingly observable. The larger concern may therefore be that NGPC fragments existing control and decreases the practicality of intervention.

\subsection{Evaluation}\label{evaluation}

Evaluating the main pathways to loss of control risk, we argue that NGPC plausibly changes several key parameters that affect the probability and severity of loss of control events. If our pathways are correct, then deployment pressure in NGPC increases the risk of the deployment of a capable yet misaligned agent in crucial contexts. Additionally, path-dependence and first-move advantage beliefs make actors more likely to take risky, irreversible steps when they believe a short time window for decision advantage exists. Due to reduced institutional capacity and implemented AI decision making, actors have an eroded ability to detect, override, and control the misaligned agent. Lastly, the existence of multiple agents and actors increase the risk of joint, multi-agent coordination and collusion failures.

Our key uncertainties include (1) whether alignment materially trades off with capability in deployment settings, (2) whether decision makers tend to prioritise stability and predictability over the capabilities of a potentially uncontrollable weapon during a conflict, (3) whether and how AI systems strengthen or weaken institutions, and (4) how prominent the perception of early military advantage powered by AI is among decision makers. Therefore, loss of control risk from NGPC seems to warrant further empirical investigation as we gain more evidence of the deployment of AI systems in NGPC.

\section{Conclusion}\label{conclusion}

This paper set out to evaluate whether NGPC is plausibly within an order of magnitude of GPC across three risk vectors: escalation to GPC (H1), catastrophic terrorism (H2), and loss of control over advanced AI (H3). Using causal models, literature reviews, and qualitative assessments, we find that an increase in the rate and severity of NGPC materially affects the key mechanisms behind each risk vector. We do not derive a ratio between NGPC and GPC risk, and the order-of-magnitude framing should be read as the decision threshold that motivated the inquiry rather than as a result. What we can say is narrower and still substantial: the null hypothesis that NGPC is a minor contributor to catastrophic risk relative to GPC is not well supported by the mechanisms we examine, and the burden of argument should shift toward those who hold it.

For GPC escalation (H1), the evidence is strongest. NGPC can draw great powers into opposing patronage relationships (Model A) and degrade system-level conditions that make GPC relatively stable (Model B). Proxy entanglement, patron realignment, and the use of conflict zones as testing grounds for new military AI provide plausible mechanisms for GPC escalation. Historical counterarguments show that GPC escalation pressure can be managed using de-escalation mechanisms, but do not imply that these mechanisms will remain effective in our model.

For catastrophic terrorism (H2), the causal pathways are clear, although empirically uncertain in magnitude. NGP conflict zones are unique environments for grasping and employing tacit operational knowledge in real-world military situations. AI may lower planning barriers, enable better coordination, and lead to capability transfer across military grounds. We therefore expect NGPC to increase some combination of the attempt rate and success rate of terrorist attacks, with a plausible but more uncertain effect on severity.

For loss of control (H3), our most speculative hypothesis, NGPC introduces mechanisms which are both concerning and possibly empirically testable in the future. NGPC may increase deployment pressure of misaligned AIs, raise incentives to grant risky affordances to misaligned AI, and create a context in which detection and mitigation of misaligned behaviours are difficult or costly. Each causal link in these pathways carries a degree of uncertainty, but the plausible conjunction of these mechanisms warrants a deeper investigation into the strength of this hypothesis, and for us, is sufficient to regard H3 as a plausible pathway to catastrophic risk from advanced AI.

Our analysis is preliminary. We do not produce causal models that estimate the degree of risk precisely, and we do not model interaction effects between pathways and intermediate variables in a fully rigorous way. Interaction effects might amplify or dampen the risk in ways our independent treatment does not consider. Additionally, there remain empirical gaps, which we have noted throughout this paper, that require more work, some of which can only be observed during NGPCs in the future.

We have set out multiple causal pathways from NGPC to catastrophic harm, assessed each against the available evidence, and found none of them negligible. That is not a demonstration that NGPC rivals GPC as a source of catastrophic risk, and we have not claimed that it is. It is a demonstration that the question is open, that the mechanisms are identifiable and in several cases already observable, and that the cause area merits the empirical and tractability work that would settle it.

\section*{Acknowledgements}
\addcontentsline{toc}{section}{Acknowledgements}

We are grateful to Liam Patell for his mentorship throughout the programme, and to many people for helpful conversations and feedback, including Charlie Alaimo, Hadley Spadaccini, Cadence James, Luke Dawes, and numerous others at ERA:AI, FIG, and GovAI for feedback and support on this work. We used LLMs to help with various parts of writing and editing this text. 

\section{Appendix}\label{appendix}

{\footnotesize\setlength{\extrarowheight}{2.5pt}\renewcommand{\arraystretch}{1.2}
\begin{longtable}[]{@{}
  >{\raggedright\arraybackslash}p{(\columnwidth - 2\tabcolsep) * \real{0.5000}}
  >{\raggedright\arraybackslash}p{(\columnwidth - 2\tabcolsep) * \real{0.5000}}@{}}
  \caption{Model A parameter definitions.}\label{tab:model-a-params}\\
\toprule\noalign{}
\begin{minipage}[t]{\linewidth}\raggedright
\textbf{Parameter}
\end{minipage} & \begin{minipage}[t]{\linewidth}\raggedright
\textbf{Definition}
\end{minipage} \\\addlinespace
\begin{minipage}[t]{\linewidth}\raggedright
NGPC Rate
\end{minipage} & \begin{minipage}[t]{\linewidth}\raggedright
Baseline frequency of NGPCs with features that make great power involvement plausible: geographic proximity to GP interests, resource salience, alliance ties, or enduring rivalry dynamics.
\end{minipage} \\\addlinespace
\begin{minipage}[t]{\linewidth}\raggedright
GP Involvement Rate
\end{minipage} & \begin{minipage}[t]{\linewidth}\raggedright
Probability that a given NGPC draws great power participation on opposing sides, whether through arms transfers, military advisors, proxy relationships, or forward deployment. Alliance activation rates range from 23-30\% against all treaty obligations (Sabrosky 1980; Siverson and King 1980) to approximately 75\% against specific military commitments (Leeds et al. 2000). This is the gateway parameter that converts a local conflict into a great power concern.
\end{minipage} \\\addlinespace
\begin{minipage}[t]{\linewidth}\raggedright
Escalation Rate
\end{minipage} & \begin{minipage}[t]{\linewidth}\raggedright
Probability that a conflict with opposing great power involvement escalates to direct GP-versus-GP confrontation through inadvertent military contact, commitment traps, crisis spirals, or alliance entrapment. Huth's (1988) analysis of 58 extended deterrence cases found a deterrence failure rate of 41.4\%.
\end{minipage} \\\addlinespace
\begin{minipage}[t]{\linewidth}\raggedright
1 - De-escalation Rate
\end{minipage} & \begin{minipage}[t]{\linewidth}\raggedright
Probability that escalation is not successfully managed through diplomacy, communication channels, off-ramps, or mutual restraint. This is distinct from the complement of escalation: a conflict can simultaneously face high escalation pressure and strong de-escalation capacity (e.g., the Cuban Missile Crisis), or face low escalation pressure with degraded de-escalation institutions.
\end{minipage} \\\addlinespace
\midrule\noalign{}
\endhead
\bottomrule\noalign{}
\endlastfoot
\end{longtable}
}

{\footnotesize\setlength{\extrarowheight}{2.5pt}\renewcommand{\arraystretch}{1.2}
\begin{longtable}[]{@{}
  >{\raggedright\arraybackslash}p{(\columnwidth - 2\tabcolsep) * \real{0.5000}}
  >{\raggedright\arraybackslash}p{(\columnwidth - 2\tabcolsep) * \real{0.5000}}@{}}
  \caption{Model B parameter definitions.}\label{tab:model-b-params}\\
\toprule\noalign{}
\begin{minipage}[t]{\linewidth}\raggedright
\textbf{Parameter}
\end{minipage} & \begin{minipage}[t]{\linewidth}\raggedright
\textbf{Definition}
\end{minipage} \\\addlinespace
\begin{minipage}[t]{\linewidth}\raggedright
NGPC Instability Effect
\end{minipage} & \begin{minipage}[t]{\linewidth}\raggedright
The degree to which ongoing NGPCs degrade the strategic environment, covering both political instability (sub-channel A) and military-technological instability (sub-channel B). Expressed as a scalar multiplier on baseline GPC risk.
\end{minipage} \\\addlinespace
\begin{minipage}[t]{\linewidth}\raggedright
Baseline GP Crisis Rate
\end{minipage} & \begin{minipage}[t]{\linewidth}\raggedright
Frequency of great power crises (diplomatic confrontations, military standoffs, brinkmanship episodes) absent NGPC effects. The ICB database documents 512 international crises from 1918-2021, with ten states triggering 30\% of all crises in the twentieth century (Brecher and Wilkenfeld 1997).
\end{minipage} \\\addlinespace
\begin{minipage}[t]{\linewidth}\raggedright
Crisis-to-GPC Escalation Rate
\end{minipage} & \begin{minipage}[t]{\linewidth}\raggedright
Probability that a given great power crisis escalates to armed conflict. Informed by the spiral model, the extended deterrence literature, and historical near-miss data (Sagan 1993). Between 1977 and 1985, NORAD held over 20,000 first-stage missile alert conferences, with 1,152 proceeding to second-stage threat evaluation (Center for Defense Information 1986).
\end{minipage} \\\addlinespace
\midrule\noalign{}
\endhead
\bottomrule\noalign{}
\endlastfoot
\end{longtable}
}

{\footnotesize\setlength{\extrarowheight}{2.5pt}\renewcommand{\arraystretch}{1.2}
\begin{longtable}[]{@{}
  >{\raggedright\arraybackslash}p{(\columnwidth - 6\tabcolsep) * \real{0.2483}}
  >{\raggedright\arraybackslash}p{(\columnwidth - 6\tabcolsep) * \real{0.2483}}
  >{\raggedright\arraybackslash}p{(\columnwidth - 6\tabcolsep) * \real{0.2500}}
  >{\raggedright\arraybackslash}p{(\columnwidth - 6\tabcolsep) * \real{0.2534}}@{}}
  \caption{Cross-cutting intermediate variables and their roles in Models A and B.}\label{tab:cross-cutting}\\
\toprule\noalign{}
\begin{minipage}[t]{\linewidth}\raggedright
\textbf{Variable}
\end{minipage} & \begin{minipage}[t]{\linewidth}\raggedright
\textbf{Role in Model A}
\end{minipage} & \begin{minipage}[t]{\linewidth}\raggedright
\textbf{Role in Model B}
\end{minipage} & \begin{minipage}[t]{\linewidth}\raggedright
\textbf{Why cross-cutting}
\end{minipage} \\\addlinespace
\midrule\noalign{}
\endfirsthead
\toprule\noalign{}
\begin{minipage}[t]{\linewidth}\raggedright
\textbf{Variable}
\end{minipage} & \begin{minipage}[t]{\linewidth}\raggedright
\textbf{Role in Model A}
\end{minipage} & \begin{minipage}[t]{\linewidth}\raggedright
\textbf{Role in Model B}
\end{minipage} & \begin{minipage}[t]{\linewidth}\raggedright
\textbf{Why cross-cutting}
\end{minipage} \\\addlinespace
\midrule\noalign{}
\endhead
\bottomrule\noalign{}
\endlastfoot
\begin{minipage}[t]{\linewidth}\raggedright
Information environment quality
\end{minipage} & \begin{minipage}[t]{\linewidth}\raggedright
Affects de-escalation capacity: degraded information environments reduce the ability to negotiate off-ramps and build trust during crises.
\end{minipage} & \begin{minipage}[t]{\linewidth}\raggedright
Affects NGPC Instability Effect (sub-channel A): AI-generated disinformation erodes trust between great powers and pollutes the information needed for diplomatic coordination.
\end{minipage} & \begin{minipage}[t]{\linewidth}\raggedright
Information quality is a precondition for both crisis management (Model A) and baseline trust maintenance (Model B). It degrades through the same mechanisms (disinformation, synthetic media) regardless of which model it feeds into.
\end{minipage} \\\addlinespace
\begin{minipage}[t]{\linewidth}\raggedright
Decision-making timeline compression
\end{minipage} & \begin{minipage}[t]{\linewidth}\raggedright
Affects escalation rate: autonomous systems and AI-assisted targeting reduce the time available for human judgment, increasing the risk of inadvertent escalation.
\end{minipage} & \begin{minipage}[t]{\linewidth}\raggedright
Affects crisis-to-GPC escalation rate: AI speed outpaces diplomatic response during great power crises.
\end{minipage} & \begin{minipage}[t]{\linewidth}\raggedright
AI-driven speed compresses both the tactical timeline within a specific conflict (Model A) and the strategic timeline during a great power crisis (Model B). The underlying driver is the same: autonomous systems operating faster than human decision-making.
\end{minipage} \\\addlinespace
\begin{minipage}[t]{\linewidth}\raggedright
Great power threat perception
\end{minipage} & \begin{minipage}[t]{\linewidth}\raggedright
Affects GP involvement rate: perceived threats from NGPC outcomes drive decisions to intervene.
\end{minipage} & \begin{minipage}[t]{\linewidth}\raggedright
Affects NGPC Instability Effect: threat inflation from NGPC outcomes raises baseline tensions between great powers.
\end{minipage} & \begin{minipage}[t]{\linewidth}\raggedright
Threat perception is shaped by the same NGPC events in both models, but operates on different timescales: acute perception drives specific intervention decisions (Model A), while chronic perception shifts the baseline security environment (Model B).
\end{minipage} \\\addlinespace
\begin{minipage}[t]{\linewidth}\raggedright
Capability diffusion
\end{minipage} & \begin{minipage}[t]{\linewidth}\raggedright
Affects NGPC rate: AI-enabled non-state actors create more dangerous NGPCs that are more likely to attract great power attention.
\end{minipage} & \begin{minipage}[t]{\linewidth}\raggedright
Affects NGPC Instability Effect (sub-channel B): technology transfer from conflict zones shifts the military balance between great powers.
\end{minipage} & \begin{minipage}[t]{\linewidth}\raggedright
The same transfer pathways (conflict zone training, state-to-proxy transfers, open-source AI diffusion) feed into both models. Capability that reaches non-state actors raises NGPC Rate (Model A); capability that reaches rival great powers raises the Instability Effect (Model B).
\end{minipage} \\\addlinespace
\begin{minipage}[t]{\linewidth}\raggedright
Norm erosion
\end{minipage} & \begin{minipage}[t]{\linewidth}\raggedright
Affects GP involvement rate and escalation rate: erosion of sovereignty norms lowers the threshold for proxy involvement; erosion of use-of-force norms increases escalation risk.
\end{minipage} & \begin{minipage}[t]{\linewidth}\raggedright
Affects NGPC Instability Effect (sub-channel A): norm degradation raises the baseline probability of crisis initiation and escalation.
\end{minipage} & \begin{minipage}[t]{\linewidth}\raggedright
Norms constrain behavior in both models. The same norm violations (autonomous weapons use in NGPC, state transfers to terrorist groups) simultaneously lower the threshold for direct involvement (Model A) and degrade the broader normative architecture that constrains great power behavior (Model B).
\end{minipage} \\\addlinespace
\end{longtable}
}

The shared intermediate variables identified above (information environment quality, decision-making timeline compression, great power threat perception, capability diffusion, and norm erosion) appear across all three pathways. These shared nodes are the highest-priority targets for further investigation, because interventions that shift them would affect multiple risk pathways at once.
\newpage
\section*{Works Cited}
\addcontentsline{toc}{section}{Works Cited}

\begingroup
\setlength{\parindent}{0pt}
\setlength{\parskip}{0.5em}
\raggedright
\small

\newcommand{\bibent}[1]{\par\hangindent=1.6em\hangafter=1 #1}

\bibent{Adams, Rose. ``Alabuga: The Latest Destination for North Korea's Drone Ambitions.'' \emph{38 North}, December 2025. \url{https://www.38north.org/2025/12/alabuga-the-latest-destination-for-north-koreas-drone-ambitions/}.}

\bibent{Altmann, J{\"u}rgen, and Frank Sauer. ``Autonomous Weapon Systems and Strategic Stability.'' \emph{Survival} 59, no. 5 (2017): 117--142.}

\bibent{Aschenbrenner, Leopold. ``Situational Awareness: The Decade Ahead.'' Self-published, June 2024.}

\bibent{Atlantic Council. ``Experts React: India and Pakistan Have Agreed to a Shaky Cease-Fire. Where Does the Region Go from Here?'' \emph{New Atlanticist}, May 2025. \url{https://www.atlanticcouncil.org/blogs/new-atlanticist/experts-react/india-pakistan-cease-fire-experts/}.}

\bibent{Bendett, Samuel, and David Kirichenko. ``Battlefield Drones and the Accelerating Autonomous Arms Race in Ukraine.'' Center for a New American Security, January 2025. \url{https://www.cnas.org/publications/commentary/battlefield-drones-and-the-accelerating-autonomous-arms-race-in-ukraine}.}

\bibent{Bengio, Yoshua, et al. \emph{International AI Safety Report 2026}. DSIT 2026/001. Department for Science, Innovation and Technology, 2026. \url{https://internationalaisafetyreport.org}.}

\bibent{Blechman, Barry M., and Douglas M. Hart. ``The Political Utility of Nuclear Weapons: The 1973 Middle East Crisis.'' \emph{International Security} 7, no. 1 (1982): 132--156.}

\bibent{Boiko, Daniil A., Robert MacKnight, Ben Kline, and Gabe Gomes. ``Autonomous Chemical Research with Large Language Models.'' \emph{Nature} 624, no. 7992 (2023): 570--578. \url{https://doi.org/10.1038/s41586-023-06792-0}.}

\bibent{Brecher, Michael, and Jonathan Wilkenfeld. \emph{A Study of Crisis}. Ann Arbor: University of Michigan Press, 2022.}

\bibent{Brent, Roger, and Greg McKelvey Jr. \emph{Contemporary Foundation AI Models Increase Biological Weapons Risk}. Santa Monica, CA: RAND Corporation, December 2025. \url{https://www.rand.org/pubs/perspectives/PEA3853-1.html}.}

\bibent{Brittain, James J. \emph{Revolutionary Social Change in Colombia: The Origin and Direction of the FARC-EP}. Pluto Press, 2010.}

\bibent{Britzky, Haley. ``US Sets Up One-Way Attack Drone Squadron in the Middle East after Reverse-Engineering Iranian Drone.'' December 2025. \url{https://www.cnn.com/2025/12/03/politics/drones-us-iran-middle-east}.}

\bibent{Broekaert, Clara, and Lucas Webber. ``AI Use in Terrorist Plots and Attacks Surges in 2025.'' \emph{Militant Wire}, December 2025. \url{https://www.militantwire.com/p/ai-use-in-terrorist-plots-and-attacks}.}

\bibent{Burdette, Zachary, Dwight Phillips, Jacob L. Heim, Edward Geist, David R. Frelinger, Chad Heitzenrater, and Karl P. Mueller. \emph{How Artificial Intelligence Could Reshape Four Essential Competitions in Future Warfare}. RR-A4316-1. Santa Monica, CA: RAND Corporation, 2026. \url{https://www.rand.org/pubs/research_reports/RRA4316-1.html}.}

\bibent{Byman, Daniel L., Chongyang Gao, Chris Meserole, and V.S. Subrahmanian. \emph{Deepfakes and International Conflict}. Washington, DC: Brookings Institution, 2023.}

\bibent{Ch{\'a}vez, Kerry, and Ori Swed. ``The Proliferation of Drones to Violent Nonstate Actors.'' \emph{Defence Studies} 21, no. 1 (2021): 1--24.}

\bibent{Chen Jian. \emph{China's Road to the Korean War: The Making of the Sino-American Confrontation}. New York: Columbia University Press, 1996.}

\bibent{Christensen, Thomas J., and Jack Snyder. ``Chain Gangs and Passed Bucks: Predicting Alliance Patterns in Multipolarity.'' \emph{International Organization} 44, no. 2 (1990): 137--168.}

\bibent{Clark, Christopher. \emph{The Sleepwalkers: How Europe Went to War in 1914}. London: Allen Lane, 2012.}

\bibent{Council on Foreign Relations. ``Territorial Disputes in the South China Sea.'' Global Conflict Tracker, 2026. \url{https://www.cfr.org/global-conflict-tracker/conflict/territorial-disputes-south-china-sea}. Accessed February 28, 2026.}

\bibent{Crenshaw, Martha. ``Transnational Jihadism and Civil Wars.'' \emph{Daedalus} 146, no. 4 (2017): 59--70.}

\bibent{Davies, Shawn, Ther{\^e}se Pettersson, Margareta Sollenberg, and Magnus {\"O}berg. ``Organized Violence 1989--2024, and the Challenges of Identifying Civilian Victims.'' \emph{Journal of Peace Research} 62, no. 4 (2025): 1223--1240.}

\bibent{Department of Homeland Security. \emph{Department of Homeland Security Report on Reducing the Risks at the Intersection of Artificial Intelligence and Chemical, Biological, Radiological, and Nuclear Threats}. Washington, DC: Department of Homeland Security, 2024. \url{https://www.dhs.gov/sites/default/files/2024-06/24_0620_cwmd-dhs-cbrn-ai-eo-report-04262024-public-release.pdf}.}

\bibent{Diehl, Paul F., and Gary Goertz. \emph{War and Peace in International Rivalry}. Ann Arbor: University of Michigan Press, 2001.}

\bibent{Drexel, Bill, and Caleb Withers. \emph{AI and the Evolution of Biological National Security Risks}. Washington, DC: Center for a New American Security, 2024. \url{https://www.cnas.org/publications/reports/ai-and-the-evolution-of-biological-national-security-risks}.}

\bibent{Europol. \emph{European Union Terrorism Situation and Trend Report 2025}. The Hague: Europol, June 2025. \url{https://www.europol.europa.eu/publication-events/main-reports/european-union-terrorism-situation-and-trend-report-2025-eu-te-sat}.}

\bibent{Faragasso, Spencer. ``Russian Lancet-3 Kamikaze Drone Filled with Foreign Parts.'' 2023. \url{https://isis-online.org/isis-reports/russian-lancet-3-kamikaze-drone-filled-with-foreign-parts}.}

\bibent{Fasanotti, Federica Saini. ``Juntas and Moscow Reshaping Sahel Alliances.'' \emph{GIS Reports}, April 2026. \url{https://www.gisreportsonline.com/r/reshaping-sahel-alliances/}.}

\bibent{Feldstein, Steven. \emph{The Rise of Digital Repression: How Technology Is Reshaping Power, Politics, and Resistance}. Oxford University Press, 2021.}

\bibent{Findley, Michael G., and Joseph K. Young. ``Terrorism and Civil War: A Spatial and Temporal Approach to a Conceptual Problem.'' \emph{Perspectives on Politics} 10, no. 2 (2012): 285--305. \url{https://doi.org/10.1017/S1537592712000679}.}

\bibent{France 24. ``Ukraine Convicts Frenchman over Euro 2016 Attack Plot.'' May 2018. \url{https://www.france24.com/en/20180522-ukraine-terrorism-convicts-frenchman-over-euro-2016-football-france-attack-plot}.}

\bibent{Gaibulloev, Khusrav, James A. Piazza, and Todd Sandler. ``Are Resident Terrorist Groups Productive in Weak States?'' \emph{Kyklos} (2025). \url{https://doi.org/10.1111/kykl.12458}.}

\bibent{Gaibulloev, Khusrav, James A. Piazza, and Todd Sandler. ``Do Failed or Weak States Favor Resident Terrorist Groups' Survival?'' \emph{Journal of Conflict Resolution} 68, no. 5 (2024): 823--848. \url{https://doi.org/10.1177/00220027231183939}.}

\bibent{Garfinkel, Ben, and Allan Dafoe. ``How Does the Offense-Defense Balance Scale?'' \emph{Journal of Strategic Studies} 42, no. 6 (2019): 736--763.}

\bibent{Glaser, Charles L.. ``The Causes and Consequences of Arms Races.'' \emph{Annual Review of Political Science} 3 (2000): 251--276.}

\bibent{Gleijeses, Piero. \emph{Conflicting Missions: Havana, Washington, and Africa, 1959--1976}. Chapel Hill: University of North Carolina Press, 2002.}

\bibent{Goldfarb, Avi, and Jon R. Lindsay. ``Prediction and Judgment: Why Artificial Intelligence Increases the Importance of Humans in War.'' \emph{International Security} 46, no. 3 (2022): 7--50.}

\bibent{Graham, Robert. ``How Terrorists Use Encryption.'' \emph{CTC Sentinel} 9, no. 6 (June 2016). \url{https://ctc.westpoint.edu/how-terrorists-use-encryption/}.}

\bibent{Hegghammer, Thomas. ``Should I Stay or Should I Go? Explaining Variation in Western Jihadists' Choice between Domestic and Foreign Fighting.'' \emph{American Political Science Review} 107, no. 1 (2013): 1--15. \url{https://doi.org/10.1017/S0003055412000615}.}

\bibent{H{\"o}ller, Linus. ``Drug Cartel Operatives Snuck into Ukraine for Drone Training: Report.'' \emph{Defense News}, July 2025. \url{https://www.defensenews.com/global/the-americas/2025/07/30/drug-cartel-operatives-snuck-into-ukraine-for-drone-training-report/}.}

\bibent{Honan, Stephen. ``Drug Cartels Are Adopting Cutting-Edge Drone Technology. Here's How the US Must Adapt.'' Atlantic Council, September 2025. \url{https://www.atlanticcouncil.org/blogs/new-atlanticist/drug-cartels-are-adopting-cutting-edge-drone-technology-heres-how-the-us-must-adapt/}.}

\bibent{Horowitz, Michael C.. ``Artificial Intelligence, International Competition, and the Balance of Power.'' \emph{Texas National Security Review} 1, no. 3 (May 2018): 36--57.}

\bibent{Horowitz, Michael C.. ``Nonstate Actors and the Diffusion of Innovations: The Case of Suicide Terrorism.'' \emph{International Organization} 64, no. 1 (2010): 33--64.}

\bibent{Horowitz, Michael C.. ``When Speed Kills: Lethal Autonomous Weapon Systems, Deterrence and Stability.'' \emph{Journal of Strategic Studies} 42, no. 6 (2019): 764--788.}

\bibent{Howard, Lise Morj{\'e}, and Alexandra Stark. ``Why Civil Wars Are Lasting Longer.'' \emph{Foreign Affairs}, February 2018. \url{https://www.foreignaffairs.com/articles/syria/2018-02-27/why-civil-wars-are-lasting-longer}.}

\bibent{Huth, Paul K.. \emph{Extended Deterrence and the Prevention of War}. New Haven, CT: Yale University Press, 1988.}

\bibent{Institute for Economics and Peace. \emph{Global Terrorism Index 2024}. Sydney: Institute for Economics and Peace, 2024.}

\bibent{Jervis, Robert. ``Cooperation Under the Security Dilemma.'' \emph{World Politics} 30, no. 2 (1978): 167--214.}

\bibent{Johnson, James. ``Artificial Intelligence, Drone Swarming and Escalation Risks in Future Warfare.'' \emph{The RUSI Journal} 165, no. 2 (2020): 26--36. \url{https://doi.org/10.1080/03071847.2020.1752026}.}

\bibent{Johnson, James. ``Inadvertent Escalation in the Age of Intelligence Machines: A New Model for Nuclear Risk in the Digital Age.'' \emph{European Journal of International Security} 7, no. 3 (2022): 337--359.}

\bibent{Kallenborn, Zachary, and Philipp C. Bleek. ``Swarming Destruction: Drone Swarms and Chemical, Biological, Radiological, and Nuclear Weapons.'' \emph{The Nonproliferation Review} 25, no. 5--6 (2018): 523--543. \url{https://doi.org/10.1080/10736700.2018.1546902}.}

\bibent{Karčić, Hikmet. ``The Balkan Connection: Foreign Fighters and the Far Right in Ukraine.'' May 2020. \url{https://newlinesinstitute.org/nonstate-actors/the-balkan-connection-foreign-fighters-and-the-far-right-in-ukraine/}.}

\bibent{Kassab, Hanna Samir. \emph{Power Vacuums and Global Politics: Areas of State and Non-state Competition in Multipolarity}. New York: Routledge, 2023.}

\bibent{Kaunert, Christian, Alex MacKenzie, and Sarah L{\'e}onard. ``Far-right Foreign Fighters and Ukraine: A Blind Spot for the European Union?'' \emph{New Journal of European Criminal Law} 14, no. 2 (2023): 247--266. \url{https://doi.org/10.1177/20322844231164089}.}

\bibent{Kirichenko, David. ``The Booming China-Russia Drone Alliance.'' Center for European Policy Analysis, June 2025. \url{https://cepa.org/article/the-booming-china-russia-drone-alliance/}.}

\bibent{Knights, Michael, and Alex Almeida. ``What Iran's Drones in Ukraine Mean for the Future of War.'' Washington Institute for Near East Policy, November 2022. \url{https://www.washingtoninstitute.org/policy-analysis/what-irans-drones-ukraine-mean-future-war}.}

\bibent{Lacina, Bethany, and Nils Petter Gleditsch. ``Monitoring Trends in Global Combat: A New Dataset of Battle Deaths.'' \emph{European Journal of Population} 21, no. 2--3 (2005): 145--166.}

\bibent{Lai, Brian. ``{\textquotesingle}Draining the Swamp{\textquotesingle}: An Empirical Examination of the Production of International Terrorism, 1968--1998.'' \emph{Conflict Management and Peace Science} 24, no. 4 (2007): 297--310. \url{https://doi.org/10.1080/07388940701643649}.}

\bibent{Landgraf, Walter, and Nareg Seferian. ``A `Frozen Conflict' Boils Over: Nagorno-Karabakh in 2023 and Future Implications.'' Foreign Policy Research Institute, January 2024. \url{https://www.fpri.org/article/2024/01/a-frozen-conflict-boils-over-nagorno-karabakh-in-2023-and-future-implications/}.}

\bibent{Leeds, Brett Ashley, Andrew G. Long, and Sara McLaughlin Mitchell. ``Reevaluating Alliance Reliability: Specific Threats, Specific Promises.'' \emph{Journal of Conflict Resolution} 44, no. 5 (2000): 686--699.}

\bibent{Levy, Jack S.. ``The Contagion of Great Power War Behavior, 1495--1975.'' \emph{American Journal of Political Science} 26, no. 3 (1982): 562--584.}

\bibent{Lewandowsky, Stephan, Werner Strizke, Alexandra M. Freund, and Klaus Oberauer. ``Misinformation, Disinformation, and Violent Conflict: From Iraq and the `War on Terror' to Future Threats to Peace.'' \emph{American Psychologist} 68, no. 7 (2013): 487--501.}

\bibent{Lin, Bonny, Brian Hart, Leon Li, Hugh Grant-Chapman, Truly Tinsley, and Feifei Hung. ``CRINK Security Ties: Growing Cooperation, Anchored by China and Russia.'' Center for Strategic and International Studies, September 2025. \url{https://www.csis.org/analysis/crink-security-ties-growing-cooperation-anchored-china-and-russia}.}

\bibent{MacDonald, Alistair. ``AI-Powered Drone Swarms Have Now Entered the Battlefield.'' \emph{Wall Street Journal}, September 2025. \url{https://www.wsj.com/world/ai-powered-drone-swarms-have-now-entered-the-battlefield-2cab0f05}.}

\bibent{Mazzucco, Leonardo Jacopo Maria. ``How the Houthis' Strikes on US MQ-9 Reaper Drones Serve a Wider Regional Agenda.'' \emph{MENASource}, February 2025. \url{https://www.atlanticcouncil.org/blogs/menasource/houthi-strikes-on-us-mq9-reaper-drones/}.}

\bibent{McDonald, Broderick. ``The Drones of Hayat Tahrir al-Sham.'' Global Network on Extremism and Technology, December 2024. \url{https://gnet-research.org/2024/12/20/the-drones-of-hayat-tahrir-al-sham-the-development-and-use-of-uas-in-syria/}.}

\bibent{McKeown, Ryder. ``Norm Regress: US Revisionism and the Slow Death of the Torture Norm.'' \emph{International Relations} 23, no. 1 (2009): 5--25.}

\bibent{Milla, Mirra Noor, Joevarian Hudiyana, Wahyu Cahyono, and Hamdi Muluk. ``Is the Role of Ideologists Central in Terrorist Networks? A Social Network Analysis of Indonesian Terrorist Groups.'' \emph{Frontiers in Psychology} 11 (2020): 333. \url{https://doi.org/10.3389/fpsyg.2020.00333}.}

\bibent{Mitre, Jim, and Joel B. Predd. \emph{Artificial General Intelligence's Five Hard National Security Problems}. Santa Monica, CA: RAND Corporation, 2025. \url{https://www.rand.org/pubs/perspectives/PEA3691-4.html}.}

\bibent{Moghadam, Assaf. \emph{Nexus of Global Jihad: Understanding Cooperation among Terrorist Actors}. New York: Columbia University Press, 2017.}

\bibent{Moghadam, Assaf. ``Terrorist Affiliations in Context: A Typology of Terrorist Inter-Group Cooperation.'' \emph{CTC Sentinel} 8, no. 3 (March 2015). \url{https://ctc.westpoint.edu/terrorist-affiliations-in-context-a-typology-of-terrorist-inter-group-cooperation/}.}

\bibent{Moloney, Ed. \emph{A Secret History of the IRA}. New York: W.W. Norton, 2002.}

\bibent{Mouton, Christopher A., Caleb Lucas, and Ella Guest. \emph{The Operational Risks of AI in Large-Scale Biological Attacks: Results of a Red-Team Study}. RR-A2977-2. Santa Monica, CA: RAND Corporation, January 2024.}

\bibent{Mullins, Sam. ``The Role of Non-State Actors as Proxies in Irregular Warfare and Malign State Influence.'' Irregular Warfare Center, December 2024. \url{https://irregularwarfarecenter.org/publications/research-studies/the-role-of-non-state-actors-as-proxies-in-irregular-warfare-and-malign-state-influence/}.}

\bibent{National Academies of Sciences, Engineering, and Medicine. \emph{Nuclear Terrorism: Assessment of U.S. Strategies to Prevent, Counter, and Respond to Weapons of Mass Destruction}. Washington, DC: National Academies Press, 2024. \url{https://nap.nationalacademies.org/catalog/27215/}.}

\bibent{National Security Agency. \emph{Content Credentials: Strengthening Multimedia Integrity in the Generative AI Era}. Cybersecurity Information Sheet. National Security Agency, January 2025. \url{https://media.defense.gov/2025/Jan/29/2003634788/-1/-1/0/CSI-CONTENT-CREDENTIALS.PDF}.}

\bibent{Oliynyk, Tetyana. ``Ukrainian Troops Detect 12,000 Targets Weekly with Help of AI.'' September 2024. \url{https://www.pravda.com.ua/eng/news/2024/09/23/7476410/}.}

\bibent{OpenAI. ``OpenAI and Hugging Face Partner to Address Security Incident during Model Evaluation.'' July 2026. \url{https://openai.com/index/hugging-face-model-evaluation-security-incident/}.}

\bibent{Panke, Diana, and Ulrich Petersohn. ``Why International Norms Disappear Sometimes.'' \emph{European Journal of International Relations} 18, no. 4 (2012): 719--742.}

\bibent{Pape, Robert. \emph{Dying to Win: The Strategic Logic of Suicide Terrorism}. New York: Random House, 2005.}

\bibent{Parada, Carolina. ``Gemini Robotics 1.5 Brings AI Agents into the Physical World.'' September 2025. \url{https://deepmind.google/blog/gemini-robotics-15-brings-ai-agents-into-the-physical-world/}.}

\bibent{Patrick, Stewart. ``Rules of Order: Assessing the State of Global Governance.'' Carnegie Endowment for International Peace, September 2023. \url{https://carnegieendowment.org/2023/09/12/rules-of-order-assessing-state-of-global-governance-pub-90517}.}

\bibent{Piazza, James A.. ``Fake News: The Effects of Social Media Disinformation on Domestic Terrorism.'' \emph{Dynamics of Asymmetric Conflict} 15, no. 1 (2021): 55--77.}

\bibent{Piazza, James A.. ``Incubators of Terror: Do Failed and Failing States Promote Transnational Terrorism?'' \emph{International Studies Quarterly} 52, no. 3 (2008): 469--488. \url{https://doi.org/10.1111/j.1468-2478.2008.00511.x}.}

\bibent{Piazza, James A.. ``Is Islamist Terrorism More Dangerous? An Empirical Study of Group Ideology, Organization, and Goal Structure.'' \emph{Terrorism and Political Violence} 21, no. 1 (2009): 62--88.}

\bibent{Rassler, Don, and Yannick Veilleux-Lepage. ``On the Horizon: The Ukraine War and the Evolving Threat of Drone Terrorism.'' March 2025. \url{https://ctc.westpoint.edu/on-the-horizon-the-ukraine-war-and-the-evolving-threat-of-drone-terrorism/}.}

\bibent{Rider, Toby J., Michael G. Findley, and Paul F. Diehl. ``Just Part of the Game? Arms Races, Rivalry, and War.'' \emph{Journal of Peace Research} 48, no. 1 (2011): 85--100.}

\bibent{R{\k{e}}kawek, Kacper. \emph{Foreign Fighters in Ukraine}. London: Routledge, 2023.}

\bibent{Robinson, Kali. ``Iran's Regional Armed Network.'' Council on Foreign Relations, April 2024. \url{https://www.cfr.org/articles/irans-regional-armed-network}.}

\bibent{Rustad, Siri Aas. \emph{Conflict Trends: A Global Overview, 1946--2024}. PRIO Paper. Oslo: Peace Research Institute Oslo, 2026.}

\bibent{Ryd{\'e}n, Pernilla, Sanni Laine, Susanna Ahlfors, Benoit Pylyser, Jonas Alberoth, Jennifer Schmidt, and Fanny Wellen. ``Tackling Mis- and Disinformation: Seven Insights for UN Peace Operations.'' SIPRI, 2023. \url{https://www.sipri.org/commentary/blog/2023/tackling-mis-and-disinformation-seven-insights-un-peace-operations}.}

\bibent{Sabbagh, Dan. ``Russia Can Keep Fighting Ukraine War throughout 2026, Says Military Thinktank.'' \emph{The Guardian}, February 2026. \url{https://www.theguardian.com/world/2026/feb/24/russia-fighting-ukraine-war-throughout-2026-military-thinktank}.}

\bibent{Sabrosky, Alan Ned. ``Interstate Alliances: Their Reliability and the Expansion of War.'' In \emph{The Correlates of War II: Testing Some Realpolitik Models}, edited by J. David Singer, 161--198. New York: Free Press, 1980.}

\bibent{Sagan, Scott D.. \emph{The Limits of Safety: Organizations, Accidents, and Nuclear Weapons}. Princeton, NJ: Princeton University Press, 1993.}

\bibent{Salih, Zeinab Mohammed. ``Conflict in Sudan: A Map of Regional and International Actors.'' Wilson Center, December 2024. \url{https://www.wilsoncenter.org/article/conflict-sudan-map-regional-and-international-actors}.}

\bibent{Scharre, Paul. ``Debunking the AI Arms Race Theory.'' \emph{Texas National Security Review} 4, no. 3 (2021): 121--132.}

\bibent{Sekeris, Petros G.. ``Propaganda and Conflict.'' \emph{Games and Economic Behavior} 153 (2025): 569--585.}

\bibent{Senese, Paul D., and John A. Vasquez. \emph{The Steps to War: An Empirical Study}. Princeton, NJ: Princeton University Press, 2008.}

\bibent{Singer, J. David, and Melvin Small. ``Correlates of War Project: International and Civil War Data, 1816--1992.'' Inter-university Consortium for Political and Social Research, January 2006. \url{https://doi.org/10.3886/ICPSR09905.v1}.}

\bibent{Siverson, Randolph M., and Joel King. ``Attributes of National Alliance Membership and War Participation, 1815--1965.'' \emph{American Journal of Political Science} 24, no. 1 (1980): 1--15.}

\bibent{Siverson, Randolph M., and Harvey Starr. \emph{The Diffusion of War: A Study of Opportunity and Willingness}. Ann Arbor: University of Michigan Press, 1991.}

\bibent{Snyder, Glenn H.. ``The Security Dilemma in Alliance Politics.'' \emph{World Politics} 36, no. 4 (1984): 461--495.}

\bibent{Staunton, Denis. ``Could Sudan's Offer of a Naval Base to Russia Spark an Effort to End the Civil War?'' \emph{Irish Times}, December 2025. \url{https://www.irishtimes.com/world/2025/12/03/could-sudans-offer-of-a-naval-base-to-russia-spark-an-effort-to-end-the-civil-war/}.}

\bibent{Theohary, Catherine A., and Kelley M. Sayler. \emph{Agentic Artificial Intelligence and Cyberattacks}. IF13151. Congressional Research Service, January 2026. \url{https://www.congress.gov/crs-product/IF13151}. Updated July 6, 2026.}

\bibent{Tusalem, Rollin F.. ``Democracies, Autocracies, and Political Stability.'' \emph{International Social Science Review} 90, no. 1 (2015).}

\bibent{UK AI Security Institute. \emph{Security Incident INC-2026-07-28-01}. AI Security Institute, 2026.}

\bibent{UK Government. ``Explaining Uncertainty in UK Intelligence Assessment.'' GOV.UK, \url{https://www.gov.uk/government/publications/explaining-uncertainty-in-uk-intelligence-assessment/explaining-uncertainty-in-uk-intelligence-assessment}.}

\bibent{United Nations Office of Counter-Terrorism, and United Nations Interregional Crime and Justice Research Institute. \emph{Algorithms and Terrorism: The Malicious Use of Artificial Intelligence for Terrorist Purposes}. New York: United Nations, 2021.}

\bibent{Uppsala Conflict Data Program. ``UCDP Definitions.'' Uppsala University, Accessed August 12, 2026.}

\bibent{Vasquez, John A.. \emph{Contagion and War: Lessons from the First World War}. Cambridge: Cambridge University Press, 2018.}

\bibent{Vasquez, John A., and Douglas M. Gibler. ``The Steps to War in Asia, 1931--45.'' \emph{Security Studies} 10, no. 3 (2001): 1--45.}

\bibent{Walden, Kemba Eneas, Chua Kuan Seah, and Dawn Song. ``How We Enhance Cybersecurity Defences before the Attackers in an AGI World.'' October 2025. \url{https://www.weforum.org/stories/2025/10/how-we-enhance-cybersecurity-defences-before-the-attackers-in-an-agi-world/}.}

\bibent{Wallace, Michael D., Brian L. Crissey, and Linn I. Sennott. ``Accidental Nuclear War: A Rising Risk?'' \emph{The Defense Monitor} 15, no. 7 (1986).}

\bibent{Washington Post. ``French suspect arrested for alleged attack plot to protest Muslim migrant surge.'' June 2016. \url{https://www.washingtonpost.com/world/french-suspect-arrested-for-alleged-attack-plot-during-soccer-tournament-ukraine-says/2016/06/06/199d7c73-9546-421c-89af-1a1d59989fc1_story.html}.}

\bibent{Weeks, Jessica L.. ``Autocratic Audience Costs: Regime Type and Signaling Resolve.'' \emph{International Organization} 62, no. 1 (2008): 35--64.}

\bibent{Weeks, Jessica L.. ``Strongmen and Straw Men: Authoritarian Regimes and the Initiation of International Conflict.'' \emph{American Political Science Review} 106, no. 2 (2012): 326--347.}

\bibent{Weimann, Gabriel, Alexander T. Pack, Rachel Sulciner, Joelle Scheinin, Gal Rapaport, and David Diaz. ``Generating Terror: The Risks of Generative AI Exploitation.'' \emph{CTC Sentinel} 17, no. 1 (January 2024): 17--24. \url{https://ctc.westpoint.edu/generating-terror-the-risks-of-generative-ai-exploitation/}.}

\bibent{Wells, David. ``Mapping Terrorist AI Use: Identifying Factors Behind a Relatively Slow Adoption Rate.'' Global Network on Extremism and Technology, September 2025. \url{https://gnet-research.org/2025/09/17/mapping-terrorist-ai-use-identifying-factors-behind-a-relatively-slow-adoption-rate/}.}

\bibent{Westad, Odd Arne. \emph{The Global Cold War: Third World Interventions and the Making of Our Times}. Cambridge: Cambridge University Press, 2005.}

\bibent{Ziemer, Henry. ``Illicit Innovation: Latin America Is Not Prepared to Fight Criminal Drones.'' Center for Strategic and International Studies, June 2025. \url{https://www.csis.org/analysis/illicit-innovation-latin-america-not-prepared-fight-criminal-drones}.}

\endgroup

\end{document}